\documentclass[preprint,12pt,authoryear]{elsarticle}

\usepackage{amssymb}
\usepackage{amsmath}

\usepackage[utf8]{inputenc}
\usepackage[english]{babel}
\usepackage{graphicx}
\usepackage{booktabs}
\usepackage{multirow}
\usepackage{array}
\usepackage{xcolor}
\usepackage{hyperref}
\usepackage{cleveref}
\usepackage{listings}
\usepackage{caption}
\usepackage{subcaption}
\usepackage{enumitem}
\usepackage{longtable}
\usepackage[table]{xcolor}

\usepackage[margin=1in]{geometry}
\hypersetup{
    colorlinks=true,
    linkcolor=blue,
    citecolor=blue,
    urlcolor=blue
}

\newcolumntype{P}[1]{>{\raggedright\arraybackslash}p{#1}}

\journal{Computers \And Education}

\begin{document}

\begin{frontmatter}



\title{CogTax: A Four-Level Cognitive Taxonomy for Command-Line Computing Education} 

\author[inst1,inst2]{Manuel Alonso-Carracedo\corref{cor1}} 
\ead{manuel.alonso.carracedo@uvigo.gal} \cortext[cor1]{Corresponding author.}
\author[inst1,inst2]{Ruben Fernandez-Boullon} 
\ead{ruben.fernandez.boullon@uvigo.gal}
\author[inst1,inst2]{Pedro Celard} 
\ead{pedro.celard.perez@uvigo.gal}
\author[inst1,inst2]{Francisco J. Rodríguez-Martínez} 
\ead{ franjrm@uvigo.gal}
\author[inst1,inst2]{Lorena Otero-Cerdeira} 
\ead{locerdeira@uvigo.gal}

\affiliation[inst1]{organization={Computer Science Department, IFCAE-Instituto de Investigación en Física, Computación y Ciencia Aeroespacial},
addressline={Universidade de Vigo}, 
city={Ourense}, 
postcode={32004}, 
country={Spain}}

\affiliation[inst2]{organization={Universidade de Vigo, Department of Computer Science, ESEI-Escuela Superior de Ingeniería Informática}, 
addressline={Edificio Politécnico, Campus Universitario As Lagoas s/n}, 
city={Ourense}, 
postcode={32004}, 
country={Spain}}

\begin{abstract}
As computing education expands beyond traditional programming into operational domains such as systems administration and command-line environments, existing pedagogical frameworks struggle to capture a dimension that is critical in these contexts: the real-world consequences of learner actions. Existing cognitive taxonomies classify learning objectives by mental operations but do not account for system impact, leaving a critical gap in command-line education where conceptually simple commands can have severe consequences. This work presents CogTax, a four-level cognitive taxonomy that integrates two dimensions: cognitive complexity, derived from Bloom's Revised Taxonomy, and operational impact, which distinguishes observational, reversible, structural, and administrative operations. The four progressive levels range from safe read-only inspection to advanced system management requiring integration of multiple abstract models. Then, the taxonomy level is defined as the maximum of these dimensions, ensuring that both conceptual understanding and operational awareness are addressed. CogTax gives instructors a principled framework for sequencing course material and calibrating assessment difficulty, and gives students an explicit reference for self-assessment and gap identification. To demonstrate that taxonomy levels are automatically assignable, making the framework scalable without manual expert annotation, a classifier that combines syntactic representations derived from abstract syntax trees with semantic embeddings is trained. Evaluated on 585 expert-annotated Linux/bash commands, this combined approach achieves 89\% accuracy, outperforming either representation alone, and demonstrates cross-language extensibility through structural equivalences across command languages.

\end{abstract}




\end{frontmatter}

\section{Introduction}

Teaching command-line computing presents a distinctive pedagogical challenge: instructors must simultaneously balance the mental operations required to understand a command (cognitive complexity) and the potential consequences of executing it (operational risk). In courses covering system administration, Linux/bash scripting, database query languages, or network configuration, students encounter commands ranging from safe read-only queries to operations that can irreversibly alter system state. Without a principled framework for organizing this complexity, instructors face three related difficulties. First, sequencing course material becomes ad-hoc, potentially introducing high-impact operations before foundational concepts are consolidated, imposing unnecessary cognitive load on learners \citep{Sweller1994}. Second, designing examinations with calibrated difficulty levels, particularly when constructing parallel versions for different groups, relies on subjective judgment rather than explicit criteria. Third, students lack an objective reference for self-assessment, making it difficult to identify competence boundaries or trace prerequisite gaps systematically \citep{Hazzan2020}.

Addressing these challenges require specific taxonomies that provide an adequate framework for command-line operations. Bloom's Revised Taxonomy and its computing-specific adaptations \citep{anderson2001,Bamkole2023} provide robust frameworks for classifying learning objectives across cognitive levels, and these have been successfully applied to programming tasks such as algorithm design and software development \citep{Geissler2023}. However, they focus primarily on tasks that produce artefacts, programs, functions, or classes, and address cognitive depth without accounting for operational impact. The reviewed literature does not identify any existing taxonomy that systematically classifies individual computing commands according to the cognitive demand they impose \textit{and} the reversibility and scope of their system effects \citep{Masapanta2018,Imbulpitiya2021}.

To address this gap, the present work introduces CogTax, a four-level cognitive taxonomy explicitly designed for command-line computing education. The taxonomy integrates two dimensions. The first one is the cognitive complexity, derived from Bloom's cognitive levels, and the second one is the operational impact, which characterizes whether a command is observational, reversible, structural, or administrative. The taxonomy level assigned to a specific material or command ensures that both dimensions are pedagogically addressed, as conceptual mastery alone is insufficient if operational awareness is absent, and vice versa. This two-dimensional structure provides instructors with a principled tool for curriculum sequencing and exam calibration, while giving students an explicit complexity reference for self-regulated learning.

While manual expert classification establishes the pedagogical validity of the taxonomy, computational approaches offer complementary advantages. This approaches enable consistent classification at scale, reduce subjective variability, and can reveal whether the taxonomy levels correspond to measurable structural and semantic properties intrinsic to command syntax. An automatic classification methodology is developed that integrates structural analysis of command syntax with semantic representations that abstract beyond surface form. The results obtained confirm that taxonomy levels can be automatically assigned with high accuracy, supporting both the internal consistency of the proposed framework and its potential applicability to broader computer science education contexts.

The remainder of this paper is organized as follows. Section~\ref{sec:related} reviews related work on cognitive taxonomies in computing education and automated classification approaches. Section~\ref{sec:methodology} presents the methodological foundations of the study, including the development of the taxonomy and the approach used for its evaluation. Section~\ref{sec:results} presents complete-dataset cross-validation results and the decision-level maximum rule. Section~\ref{sec:discussion} discusses pedagogical implications, assessment calibration, and the cross-language generalization pathway. Finally, Section~\ref{sec:conclusion} summarizes the main contributions and outlines directions for future work.

\section{Related Work}
\label{sec:related}

Cognitive taxonomies provide structured frameworks for organizing learning objectives and instructional content by complexity level \citep{anderson2001}. While Bloom's Revised Taxonomy and its computing-specific adaptations have been successfully applied to programming tasks, their application to individual computing commands as pedagogical units remains underexplored. The development of a command-level cognitive taxonomy requires grounding in three complementary research areas: (i) learning taxonomies in computing education, which provide the conceptual foundation for classifying cognitive demand; (ii) theories of cognitive load and constructivist scaffolding, which inform how complexity gradients support learning progression; and (iii) automated classification approaches, which demonstrate the technical feasibility of operationalizing taxonomic distinctions computationally. This section reviews each area and positions the proposed taxonomy within this landscape.

\subsection{Learning taxonomies in computing education}

Bloom's Revised Taxonomy \citep{anderson2001} organizes learning objectives into six cognitive levels: Remember, Understand, Apply, Analyze, Evaluate, and Create. It has been widely adopted in computing education as a framework for aligning instruction with assessment \citep{Bamkole2023}. Gaber et al. \citep{Gaber2023} analyzed exam questions using Bloom's levels and found that most questions clustered at the Understand level, suggesting systematic underexposure to higher-order tasks. Tang et al.\ \citep{Tang2024} proposed explicit mappings from computing task types to Bloom's levels, observing that design and algorithm synthesis tasks consistently engage higher-order cognitive skills. 

Curriculum-level frameworks such as Computing Curricula 2020 (CC2020) \citep{CC2020} and Computer Science Curricula 2023 (CSC23) \citep{Kumar2023} represent the most authoritative attempts to organise computing knowledge and skills at scale. CC2020 moved away from knowledge-based learning toward competency-based learning, defining competency as the combination of knowledge, skills, and dispositions within a task context. CSC23 extended this model with a revised knowledge framework and a flexible competency structure adaptable to individual programme goals. While both frameworks are comprehensive in their coverage of computing disciplines (e.g. systems, software engineering, or data science) they operate at the granularity of knowledge areas and programme-level outcomes rather than individual tasks or commands. Neither assigns cognitive levels to specific tasks, nor do they address the operational impact of commands within interactive computing environments. They therefore provide a useful backdrop for situating the present work but do not offer the fine-grained, task-level classification that CogTax targets. 

Within the narrower domain of query languages, recent work has proposed structural taxonomies of SQL complexity aimed at benchmarking and evaluation. For instance, existing text-to-SQL datasets are focused on read operations and propose a taxonomy that systematically covers the full spectrum of SQL query types and structural complexity, from simple projections to nested subqueries and multi-table operations \citep{wang2025}. While this represents a meaningful step toward classifying command-level complexity in an operational domain, it is motivated by Natural Language Processing (NLP) benchmarking rather than educational design. Cognitive level is not part of the classification scheme, and the framework offers no guidance on sequencing queries for instruction, assessing learner readiness, or mapping query complexity to learning outcomes.

These applications demonstrate the utility of taxonomic frameworks for programming artifacts or curriculum competency frameworks. These frameworks themselves do not account for operational risk and system impact, which are dimensions particularly relevant to command-line computing education where a conceptually simple command may have severe consequences. The reviewed literature does not identify any existing taxonomy that jointly models these two dimensions for command-based computing domains.

While taxonomies provide structural frameworks for organizing learning objectives by cognitive level, they do not explain how learners actually process new material or why some tasks are more demanding than others. Cognitive Load Theory addresses this gap by providing a theoretical foundation for understanding the cognitive constraints learners face and how these constraints influence the design of effective instruction.

\subsection{Cognitive Load Theory and Constructivist Frameworks in Computing Education}

Cognitive Load Theory (CLT) \citep{Sweller1994,Sweller2011} distinguishes three types of cognitive load: intrinsic load (inherent task complexity), extraneous load (complexity introduced by poor instruction or interface design), and germane load (productive cognitive effort directed toward schema formation). Programming tasks are particularly demanding because they require simultaneous tracking of syntax, semantics, control flow, and data state, generating high intrinsic load even for relatively simple programs.

\citet{Chandler1991} showed that presenting worked examples reduces extraneous cognitive load, making more mental capacity available for building lasting knowledge structures. This supports ordering instructional content from lower to higher intrinsic complexity, so that learners are not overloaded before foundational understanding is established. \citet{Duran2022} studied CLT in code tracing tasks and documented the relationship between working memory capacity and programming performance. While CLT provides theoretical grounding for complexity-based curriculum sequencing, it does not prescribe how to measure intrinsic load for specific programming constructs or commands, particularly when operational consequences must be considered alongside cognitive demands. This measurement challenge becomes especially acute in systems programming contexts, where effective scaffolding depends on operationalizing the cognitive load construct.

The need for explicit complexity measures extends beyond cognitive load management, as constructivist pedagogy also depends on them to determine where instructional material falls relative to each learner's current competence. Constructivism \citep{Vygotsky1978,Piaget1952} holds that learners build knowledge through active engagement with material that lies within their Zone of Proximal Development (ZPD). These are tasks that are just beyond current competence but achievable with appropriate support. In computing education, constructivist principles underlie problem-based learning \citep{Ji2025}, pair programming \citep{Hawlitschek2023}, and exploratory programming micro-worlds \citep{Levin2025}. Effective constructivist scaffolding requires an operationalized complexity measure to reliably place material in the productive challenge zone and enable students to identify their own ZPD boundaries.

Universal Design for Learning \citep{Redstone2024} complements constructivism by emphasizing multiple means of engagement, representation, and expression. Together, these frameworks suggest that effective computing education requires explicit complexity gradients that support diverse learning pathways while maintaining clear progression from foundational to advanced competencies. A taxonomy with explicit ordered levels, as the proposed in this work,  directly serves this need by giving instructors a principled basis for sequencing material by complexity, and gives students a concrete reference for identifying their current level and recognizing when the conditions for advancing to the next have been met. 

\subsection{Automated Classification and Categorization in Computing Education}

Operationalizing taxonomy-based complexity gradients requires command-level metrics that scale beyond manual expert annotation, making automated classification a practical requirement. Automatically classifying code artefacts and programming tasks by difficulty, cognitive level, or conceptual category has been an active line of research in computing education \citep{Masapanta2018}. Early work focused on labelling exercise difficulty post-hoc from student performance data, while more recent approaches aim to predict difficulty before deployment, as an example,  \citet{Wang2024} automatically classify it using the text and solution. At the task level, \citet{Kim2024} collected problem samples from Codeforces, a large competitive programming website to build a dataset that allows the training of machine learning models to automatically classify the task difficulty. \citet{Artser2024} applied clustering to student solution spaces to identify common conceptual strategies across large-scale programming courses.

A parallel line of work applies Bloom's Taxonomy directly as a classification target, assigning existing exam questions or learning activities to cognitive levels automatically. \citet{Gani2023} trained classifiers on question text to predict Bloom's level for programming assessments, and \citet{Li2022} applied similar approaches to map learning objectives in course syllabi. These efforts demonstrate that a task's cognitive level can be automatically inferred from its textual and structural properties with reasonable accuracy, without relying on manual expert annotation at scale \cite{Kumar2025}.

However, the unit of classification is almost always a programming exercise, a function, or a student-submitted program. These artefacts are evaluated for correctness, style, or conceptual depth after the fact. The classification of individual computing commands and queries as pedagogical units remains comparatively underexplored: commands differ from programs in that their cognitive and operational properties are intrinsic to their syntax rather than to the solution strategy a student employs, making them sensitive to structural classification methods that do not rely on student performance data.

The present work addresses this gap by proposing CogTax, a taxonomy in which the command itself is the unit of analysis. CogTax adapts Bloom's framework in two ways: it compresses the six cognitive levels into four, providing a scale that maps directly to observable command properties and is more actionable for course organization and student self-assessment; and it introduces an operational impact dimension absent from the reviewed taxonomic frameworks, reflecting the specific demands of systems programming.

\section{Methodology}\label{sec:methodology}

This section presents the methodological framework of the study in two complementary parts. The first formally defines CogTax, a four-level cognitive taxonomy for command-line computing education: its two constituent dimensions (cognitive complexity and operational impact) are specified and combined into a classification scheme via a maximum rule, and the curriculum sequencing principles the taxonomy enables are grounded in constructivist scaffolding theory. The second addresses computational recoverability: it describes the dataset of expert-annotated Linux/bash commands on which classifiers are trained and evaluated, and presents the two complementary command representation strategies (structural features derived from Abstract Syntax Trees and dense semantic embeddings) used to assess whether taxonomy levels are automatically assignable from command syntax alone.

\subsection{Taxonomy}\label{sec:taxonomy}

The taxonomy proposed in this work (CogTax) was developed and validated in the context of a second-year undergraduate course providing a practical introduction to Linux system administration. The course spans the full operational range of the operating system environment: basic session management and filesystem navigation, file and directory manipulation, text editing, permission and link management, data stream processing with filters and regular expressions, and process management including foreground/background execution, signal handling, and compilation. This great range of contents, from read-only information retrieval commands to multi-concept process management operations, provides a natural basis for a complex taxonomy: the commands in scope span all four defined levels, from trivial inspection tasks to operations requiring systemic understanding of the operating system.

The proposed taxonomy comprises four different categories designed following the principles of Bloom's Taxonomy. CogTax integrates two dimensions. First, \textit{cognitive complexity} $C$, that is, the mental operations required to use commands effectively (derived from Bloom's Revised Taxonomy) and second, \textit{operational impact} $O$, that is, the degree to which commands modify system state and the reversibility of those modifications. This two-dimensional approach addresses a gap in traditional taxonomies that excellently characterizes cognitive depth, but do not account for the practical risk and consequences inherent in system administration tasks. The taxonomy explicitly considers whether actions are observational (read-only, zero risk), reversible (modifiable, low risk), structural (altering logical organization, moderate risk), or  administrative (system-wide impact, high complexity).

The taxonomy level of a given command is defined by Equation~\eqref{eq:taxonomy}. This formula ensures monotone coverage: a command at level $L$ requires at least level-$L$ understanding or produces at least level-$L$ effects, but not necessarily both. Conceptual mastery alone is insufficient if operational awareness is absent and conversely.

\begin{equation}
    \label{eq:taxonomy}
    \text{L} = \max(C, O)
\end{equation}

\begin{figure}[!ht]
    \centering
    \includegraphics[width=\textwidth]{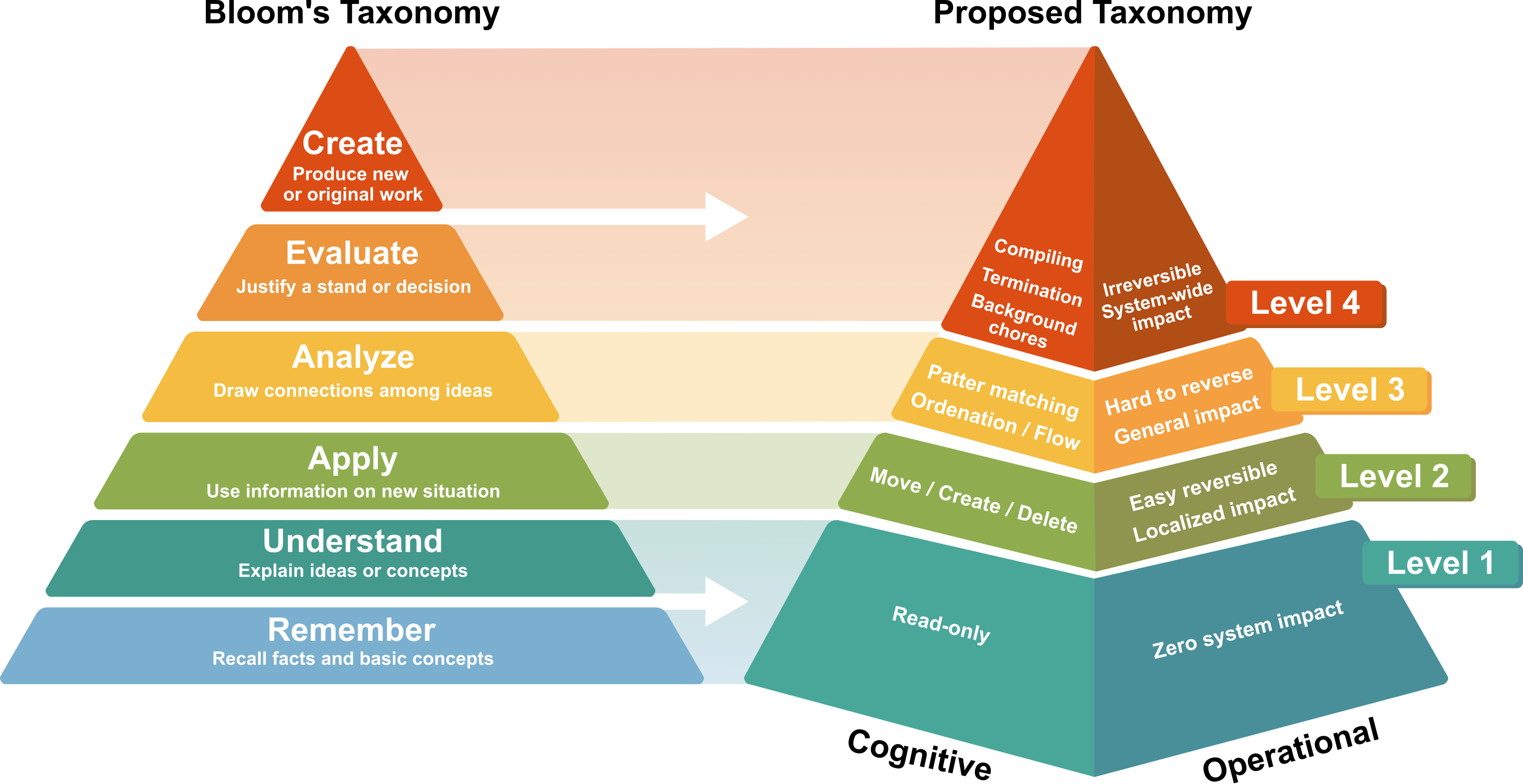}
    \caption{Proposed taxonomy compared to Bloom's taxonomy.}
    \label{fig:bloom_alignment}
\end{figure}

Figure~\ref{fig:bloom_alignment} presents the explicit alignment between CogTax and traditional Bloom's Taxonomy.

The following subsections describe each level in detail, specifying its cognitive profile, operational characteristics, and the pedagogical rationale for its placement within the progression, contextualized for the Linux/bash domain used as the empirical basis of this work.

\subsubsection{Level 1: Information Query and Observation}
This level (L1) corresponds to the first two levels of Bloom's Revised Taxonomy: Remembering and Understanding. It covers commands that retrieve and display system information without performing any modifications to the system state.

At the Remembering level, students engage in retrieving, recognizing, and recalling information from memory. This manifests when students learn to identify the meaning of command output or recall the syntax for displaying directory contents. At the Understanding level, students construct meaning from the instructional messages presented by these commands. For instance, interpreting the output of \texttt{ls -l} requires understanding what each column represents, though it does not yet require knowing how to modify those attributes.

Commands at this level impose minimal cognitive load on learners, allowing them to focus on interpreting output rather than predicting consequences of actions. Students at this stage do not need to develop a mental model of how the system maintains its state; they only need understanding of what the presented information means. This characteristic makes L1 commands ideal for safe exploration since students can experiment freely without fear of corrupting data or breaking system functionality.

The pedagogical rationale for beginning with observational commands aligns strongly with the Cognitive Load Theory~\citep{Sweller1994}, which recommends minimizing extraneous cognitive load during initial learning phases. By establishing a foundation of system literacy without the anxiety associated with potentially destructive operations, instructors can ensure that working memory resources are devoted entirely to understanding concepts rather than managing fear of consequences.

The complete list of Linux/bash commands included in this taxonomy level, along with a brief justification for their classification, is provided in Table~\ref{tab:level1_commands}.

\subsubsection{Level 2: Basic Modifications and Reversible Operations}

The second level (L2) corresponds to the Apply level of Bloom's Revised Taxonomy. At the Applying level, students use procedures in both familiar and unfamiliar situations, executing operations whose steps are known but whose context varies. This manifests in filesystem manipulation: a student who has first encountered \texttt{cp} in a guided example subsequently applies it to copy files across arbitrary directory paths or with different flag combinations, transferring the procedure without further instruction. The emphasis is on procedural fluency rather than conceptual depth, as students do not yet need to understand inode structures or filesystem internals to execute the correct sequence of operations to achieve the intended outcome.

L2 encompasses operations that create, modify, move, or delete files and directories through simple, reversible procedures that do not require understanding of underlying system models. This category includes basic file and directory manipulation operations, as well as file creation utilities and basic text editing capabilities. The cognitive profile at this level emphasizes procedural knowledge, meaning that students learn step-by-step procedures without necessarily understanding internal mechanisms. 

Learning develops through action-consequence mapping, where understanding emerges from repeated application following the pattern "if I do X, then Y happens." At this stage, schema integration remains limited, with operations learned somewhat independently rather than as parts of an integrated system model. From an operational perspective, L2 operations exhibit several important characteristics that support safe learning. Most operations can be undone: directory creation is reversed by directory removal, file compression by decompression, and archive creation by archive extraction. Effects are typically localized, confined to specific files or directories rather than affecting system-wide state. The overall risk profile remains low, as errors generally affect only the user's own files rather than system integrity or other users' data. This reversibility and localized impact create a safe practice environment where students can learn from mistakes without catastrophic consequences.

The complete list of commands included in this taxonomy level, along with a brief justification for their classification, is provided in Table~\ref{tab:level2_commands}.

\subsubsection{Level 3: Structural Understanding and Internal Models}

Level 3 (L3) corresponds to the Analyze level of Bloom's Revised Taxonomy. At the Analyzing level, students decompose information into its constituent components, discern how parts relate to each other, and construct coherent representations of underlying system organization. This manifests when students examine how permission bits compose into an access control model, trace how data flows through standard streams and redirections, or decompose a regular expression into its constituent metacharacters and quantifiers. The cognitive shift from L2 to L3 is not one of scale but of kind: students move from executing known procedures to reasoning about the structural properties and internal logic that determine why those procedures produce the effects they do.

This level represents a qualitative shift in the taxonomy, encompassing commands and operators that require understanding of abstract system models. Three primary conceptual frameworks dominate this level: the permission architecture, basic data flow mechanisms and redirection operators and pattern matching formalisms. Text processing filters also belong to this level, as their effective use requires understanding of data stream concepts together with sorting keys and filter sets.

Success at L3 requires construction of three primary mental models. The permission system model encompasses user categorization, permission types, both octal and symbolic representation systems, and the mechanics of permission inheritance and \texttt{umask} operation. Students must understand, for instance, that \texttt{chmod 754} is not merely a magical incantation but a specification that owner receives all permissions ($7 = 111_2 = \texttt{rwx}$), group receives read and execute ($5 = 101_2 = \texttt{r-x}$), and others receive only read ($4 = 100_2 = \texttt{r--}$).

The data flow model requires understanding standard streams and redirection semantics. Students must develop the ability to trace how data moves through a sequence of commands, understanding that \texttt{command > output 2 > errors} creates a complex flow where stdout of \texttt{command} goes to the file \texttt{output} while its stderr goes to \texttt{errors}.

The pattern matching model demands comprehension of regular expression syntax, greedy versus non-greedy matching, metacharacter semantics, and character classes including Portable Operating System Interface (POSIX) extensions. Students must understand that \texttt{grep '\^{}[[:alpha:]]\{3\}\$'} is not simply a search command but a formal specification requesting lines containing exactly three alphabetic characters.

This level represents the most significant pedagogical challenge in the curriculum. Students must transition from multi-structural understanding (knowing multiple independent facts) to relational understanding (integrating facts into coherent structures). This is precisely the leap required when students move from knowing that \texttt{chmod} changes permissions and \texttt{umask} sets defaults to understanding how these commands interact within a unified permission model.

The integration of multiple abstract models creates what Sweller terms ``element interactivity''~\citep{sweller1988}, the need to process multiple interrelated information elements simultaneously. Schema construction becomes the primary cognitive activity at this level. Rather than learning isolated procedures, students must build mental models of invisible system structures. 

The complete list of commands included in this taxonomy level, along with a brief justification for their classification, is provided in Table~\ref{tab:level3_commands}.

\subsubsection{Level 4: Advanced System Management and Integration}

The fourth (L4) and final level encompasses operations that require systemic understanding of operating system mechanisms, including process lifecycle, inter-process communication, network protocols or compilation toolchains. This level includes advanced process management (controlling process execution states, managing foreground and background processes, and process termination), secure remote file transfer operations, pipeline construction or source code compilation.

L4 integrates the two highest levels of Bloom's Revised Taxonomy: Evaluating and Creating. At the Evaluating level, students make judgments based on criteria and standards through checking and critiquing. This manifests when students must decide whether to use SIGTERM or SIGKILL to terminate a process, or when they must validate whether a remote connection is properly secured. At the Creating level, students put elements together to form a coherent or functional whole, reorganizing elements into new patterns or structures. 

At this level, students must develop metacognitive awareness, the ability to monitor and regulate their own problem-solving strategies. This self-regulation distinguishes L4 from lower levels where students can succeed by following prescribed procedures. Here, the integration of multiple models reaches its full complexity at this level. Understanding process management requires simultaneously reasoning about process states, signal semantics, job control, and shell behavior. Secure file transfer demands integration of file system concepts, network protocols, authentication mechanisms, and encryption principles. This multi-model integration creates cognitive complexity that exceeds simple additive combination. Understanding how concepts interact proves more demanding than understanding each concept individually.

Problem decomposition becomes essential at this level. Students face tasks that cannot be accomplished through single commands but require breaking complex goals into manageable subtasks. For instance, finding all files modified in the last week and compressing them into a dated archive requires decomposing the goal into search (find), filtering (date comparisons), and archival (tar) subtasks, then integrating these components into a coherent solution.

From an operational perspective, L4 commands exhibit characteristics that distinguish them from earlier levels. They often have system-wide impact rather than localized effects (e.g. terminating a process affects not just files but running computations). Many operations are irreversible in practical terms: once a process is killed, its in-memory state is lost; once data is transmitted over a network, it cannot be ``unsent''. 

The Bloom's for Computing framework notes that certain computing tasks require simultaneous operation at multiple cognitive levels. Debugging an error exemplifies this multi-level engagement. Students must Understand what specific errors indicate, Analyze to locate problems, Evaluate potential solutions, and Create a corrected instruction. This integration of cognitive levels distinguishes expert performance from novice approaches that remain stuck at lower levels.

At L4, students transition from following procedures to making informed decisions. This represents Bloom's distinction between lower-order thinking skills (knowledge acquisition and comprehension) and higher-order thinking skills (application of knowledge in novel contexts with evaluation and creation). A student who can only execute taught procedures remains at lower cognitive levels; a student who can evaluate the appropriateness of different approaches and create novel solutions demonstrates higher-order thinking.

The complete list of commands included in this taxonomy level, along with a brief justification for their classification, is provided in Table~\ref{tab:level4_commands}. Each command was assigned a taxonomy level by the expert annotators based on the framework presented. During the taxonomy development process, the annotators collectively discussed ambiguous cases and established consensus classification rules that were subsequently applied uniformly across the dataset. These discussions served as an implicit quality control mechanism, refining the taxonomy definitions and ensuring consistent application of the classification criteria.

\subsection{Constructivist Scaffolding Implications}

The proposed taxonomy maps directly to a constructivist curriculum sequencing principle: each level's prerequisites should be fully consolidated before the next level is introduced. In practical terms, course units that introduce $\text{L}(k{+}1)$ material should be designed assuming mastery of all L1\ldots$\text{L}(k)$ concepts. Assessments should carry explicit taxonomy-level tags, enabling post-hoc analysis of whether the difficulty distribution of an exam matches instructional intent. Worked examples should be provided at the boundary between current mastery and next level (the ZPD boundary), with the two dimensions ($C$ and $O$) separated where possible, introducing conceptually complex but operationally safe commands before operationally impactful ones.

The $L = \max(C, O)$ decomposition has a specific implication for introductory courses: \textit{operational safety should be established before operational power is introduced}. A student should develop fluency with L1--L2 operational impact commands before encountering L3--L4 commands, even if the cognitive patterns of some L3 commands (e.g., pipelines) are conceptually accessible earlier.

While the pedagogical validity of the taxonomy is established through expert consensus and grounding in Bloom's framework, manually classifying commands at each level is extremely labor-intensive work. The space of possible Linux/bash commands is combinatorially large, and expert annotation does not scale to comprehensive coverage of command variants, cross-language extensions, or dynamic curriculum adaptation. Beyond practical scalability, a more fundamental question arises: do the taxonomy levels correspond to intrinsic structural or lexical characteristics recoverable from the command syntax itself? If classification can be performed automatically with high accuracy, it would provide evidence that the taxonomy captures objective properties of command complexity rather than arbitrary expert judgment, and it would enable instructors to classify novel commands on demand without requiring domain expertise for each case. To address both the scalability challenge and the question of computational recoverability, an evaluation is conducted to determine whether taxonomy levels are automatically assignable using machine learning classifiers trained on a dataset of expert-annotated commands, described in the following section.

\subsection{Dataset}\label{subsec:dataset}

The classification evaluation requires a dataset of Linux/bash commands annotated with taxonomy levels by domain experts. To ensure that the evaluation reflects real instructional conditions and achieves comprehensive coverage of the taxonomy space, the dataset combines commands extracted from authentic student assessments with synthetically generated commands designed to achieve balanced representation across all four levels.

The complete dataset consists of 585 Linux/bash commands organized into two partitions: a held-out test partition of 117 commands and a training partition of 468 commands. The test partition was extracted from examination questions administered to second-year undergraduate students enrolled in a Computer Engineering program, specifically an Operating Systems course. These commands represent authentic instructional material spanning the full operational range of the course curriculum, from basic file inspection to multi-stage pipeline construction and process management. Using real exam commands as the test partition ensures that classifier evaluation reflects performance on material actually encountered in educational practice rather than artificially constructed examples.

The training partition comprises 468 commands synthetically generated by three computer science professors with extensive experience in systems programming instruction. To ensure comprehensive coverage of the taxonomy space and avoid distribution bias, the experts deliberately constructed commands to achieve approximate balance across the four taxonomy levels: L1 (143 commands, 24.4\%), L2 (151 commands, 25.8\%), L3 (144 commands, 24.6\%), and L4 (147 commands, 25.1\%). This balanced distribution contrasts with the naturally skewed distributions typically observed in educational contexts, where introductory-level commands predominate, and ensures that the classifier receives sufficient training signal for all taxonomy categories including the underrepresented higher levels.

The resulting 585-command dataset provides the empirical foundation for evaluating whether taxonomy levels are computationally recoverable from command syntax and semantics, as reported in the following sections.

\subsection{Commands Representation and Generalization}

Classifying a command by taxonomy level requires representing it in a form that captures the features relevant to cognitive and operational complexity. Raw command strings carry useful semantic information (e.g. the presence of a pipe, a subshell, or a permission flag is lexically visible) but they also carry a great deal of noise (e.g. specific file names, numeric arguments, and variable identifiers) that is irrelevant to taxonomy level yet dominate raw string distance. Two commands that differ only in the path they operate on should receive the same level assignment, but a string-based classifier may treat them as different.

\begin{figure}[!ht]
    \centering
    \includegraphics[width=\textwidth]{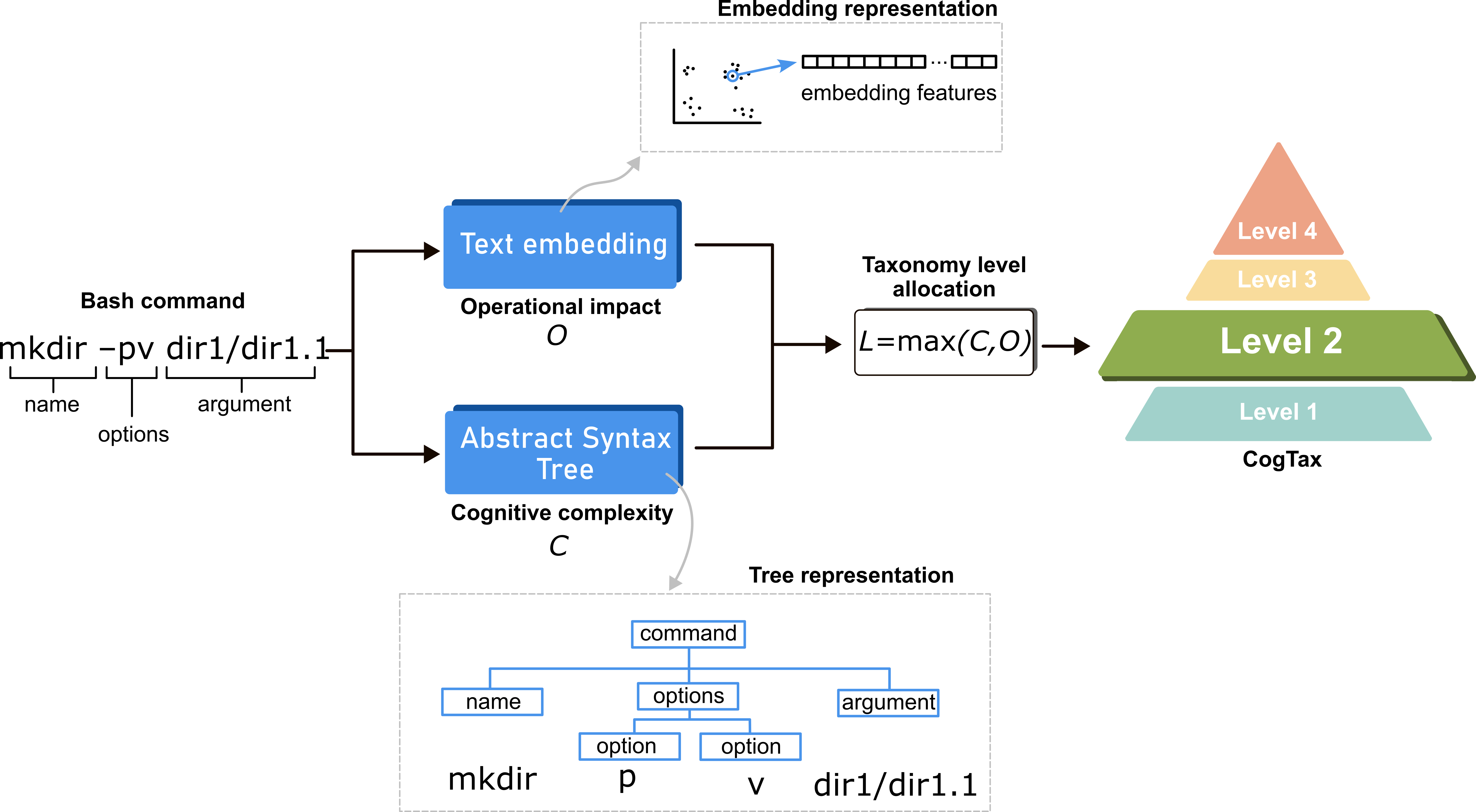}
    \caption{Overview of the classification pipeline.}
    \label{fig:general_pipeline}
\end{figure}

In order to address this problem from complementary angles, two representation approaches are explored. Structural representations parse the command into a syntax tree and extract features from its operator types, nesting depth, pipeline length and other structure characteristics, abstracting away all lexical content. Embedding-based representations encode the command as a dense vector in a semantic space where commands with similar meaning and structure are geometrically proximate, regardless of structure form. Figure~\ref{fig:general_pipeline} illustrates the overall pipeline and the following subsections describe each approach in detail.

\subsubsection{Abstract Syntax Tree Extraction}
\label{sec:ast}

An Abstract Syntax Tree (AST) is a tree-structured representation of the syntactic content of a program or command \citep{Aho2006}. Each internal node represents a syntactic construct (e.g., a pipeline, a redirection, a command substitution), and each leaf node represents a terminal token (e.g. a command name, an argument, a literal value). Unlike a raw string representation, an AST abstracts away surface-level details, retaining only the structural relationships that determine the behaviour of the command. This makes ASTs well-suited to classification tasks where the relevant signal lies in structural complexity rather than lexical content.

Linux/bash commands are parsed into ASTs using bashlex \citep{Kamara2023}, a Python parser that produces a full syntax tree for any valid Linux/bash command. The resulting AST captures: command names, options, arguments, and operator sequences; pipe (\texttt{|}), redirection (\texttt{>}, \texttt{<}), and logical (\texttt{\&\&}, \texttt{||}) operators; subshells \texttt{\$(...)} and command groups \texttt{\{...;\}}; and loops and conditionals when present. From each AST a structural feature vector is extracted consisting of tree metrics: total node count (\textit{n\_nodes}, the number of nodes across the entire tree), tree depth (\textit{depth}, the length of the longest path from root to any leaf), maximum level width (\textit{max\_width}, the largest number of nodes found at any single tree level), leaf count (\textit{n\_leaves}, the number of terminal nodes representing individual tokens), branching factor (\textit{b\_factor}, the average number of children per internal node), and node kinds count: pipelines (\textit{n\_pipelines}, the number of \texttt{|} operator nodes connecting command sequences), operators (\textit{n\_operators}, the count of logical connectors such as \texttt{\&\&} and \texttt{||}), redirections (\textit{n\_redirects}, the number of I/O redirection nodes such as \texttt{>}, \texttt{<}, and \texttt{>>}), commands (\textit{n\_commands}, the number of distinct command invocation nodes), and subshells (\textit{n\_subshells}, the count of \texttt{\$(...)} command substitution constructs).

The scalar metrics above summarize global tree properties but do not capture how syntactic constructs are locally arranged relative to one another: two commands may share identical node counts yet differ structurally in how those nodes are connected. To encode local arrangement, each command is additionally represented by the multiset of ordered pairs $(parent\_label, child\_label)$ extracted from the AST. This bag-of-edge-pairs representation records which syntactic constructs appear as direct parents of which others, providing a structural fingerprint that goes beyond aggregate counts. For example, a pipeline node whose children are a command with redirection and a command with a subshell yields a distinct signature from a pipeline whose children are two simple commands, even though both have identical pipeline and command counts.

The full AST feature vector is formed by concatenating two components. The first encodes the node-kind sequence produced by a Depth-First Search (DFS) preorder traversal of the tree: starting at the root, the algorithm visits each node before descending into its children (left to right), flattening the tree into an ordered list of syntactic type labels. For example, \texttt{cut -f1 file | grep x} yields the sequence \texttt{pipeline command word word word command word word}. $n$-grams are then extracted from this sequence: every consecutive sub-sequence of exactly $n$ labels, for $n = 1, \ldots, 4$. A unigram ($n{=}1$) records the count of a single node type; a 4-gram ($n{=}4$) such as \texttt{pipeline command word word} captures a richer structural pattern involving four consecutive constructs. TF-IDF weighting is applied to these $n$-grams, amplifying patterns that are distinctive of particular taxonomy levels and down-weighting those that appear uniformly across all commands. The second component is the ten numeric tree metrics described above (node count, depth, branching factor, etc.) after standard scaling.

\subsubsection{AST Graphic Representation}

\begin{figure}[!ht]
  \centering
  \includegraphics[width=0.75\linewidth]{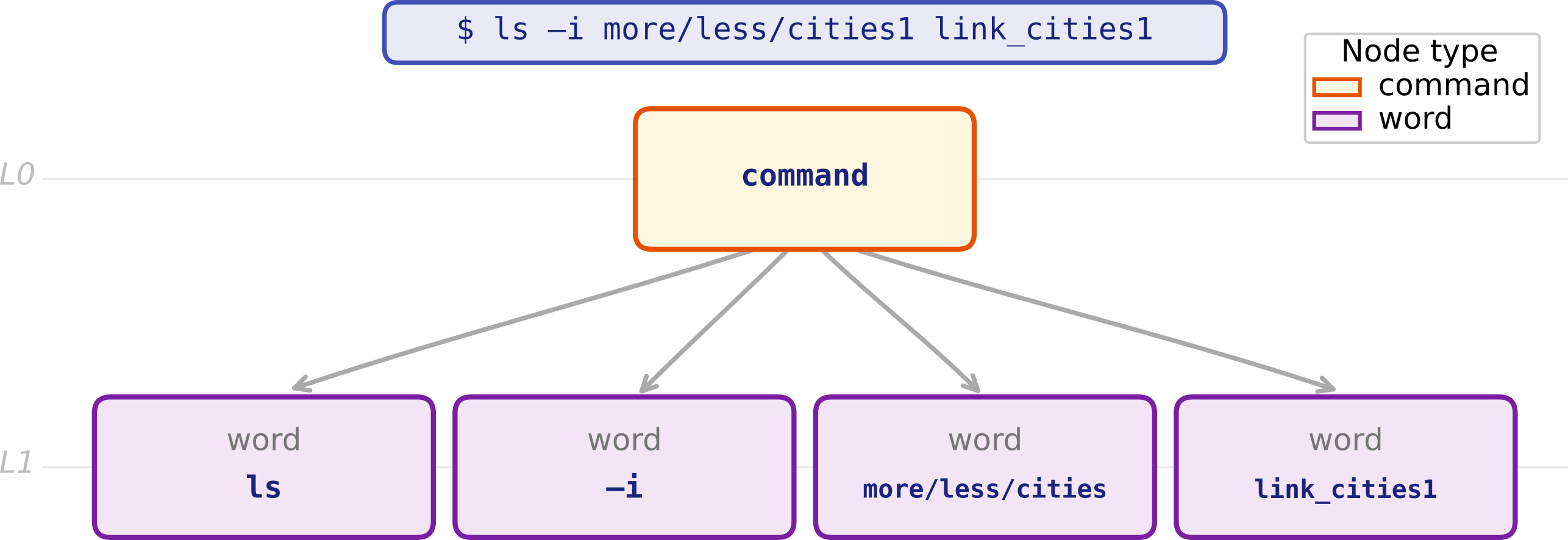}
  \caption{AST for a L1 command (\texttt{ls -{}-i more/less/cities1 link\_cities1}). A single \textit{command} node with four \textit{word} children. No operators, no redirection, and zero system impact: the command only reads directory metadata.}
  \label{fig:ast_l1}
\end{figure}

\begin{figure}[!ht]
  \centering
  \includegraphics[width=0.75\linewidth]{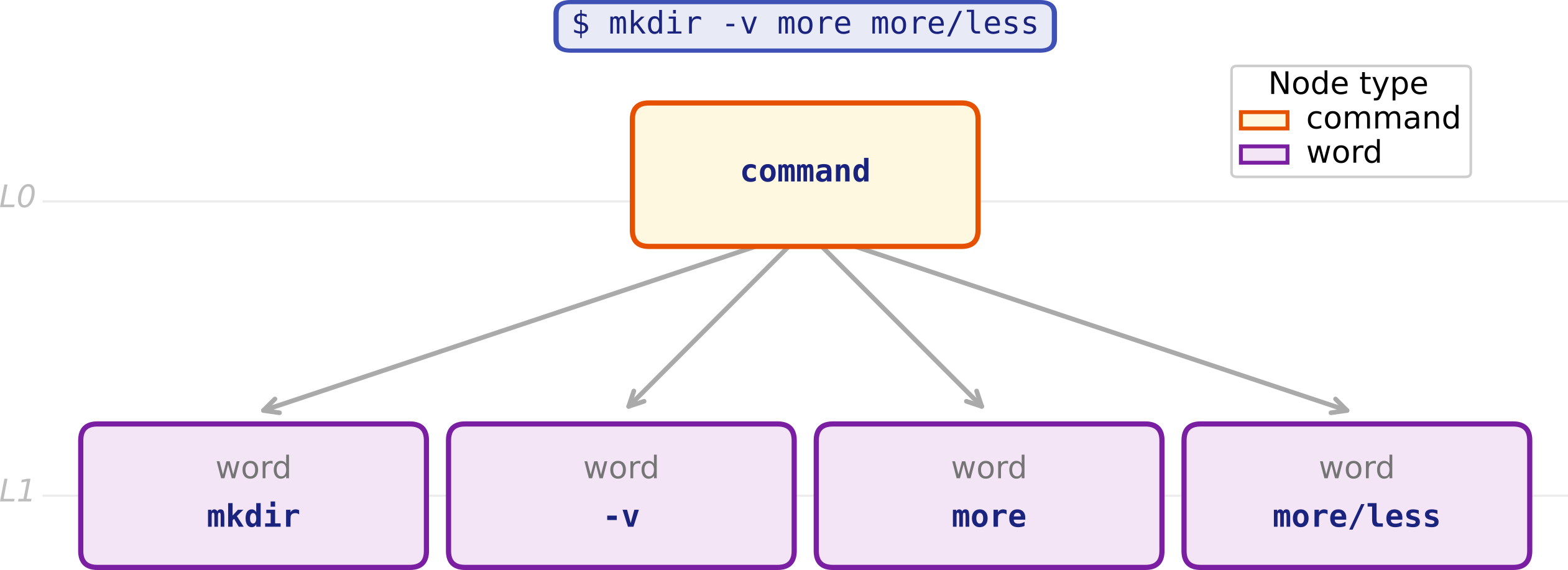}
  \caption{AST for a L2 command (\texttt{mkdir -v more more/less}). A single \textit{command} node with three \textit{word} children. The operation modifies the filesystem but is fully reversible (directories can be removed), placing it at Level~2.}
  \label{fig:ast_l2}
\end{figure}

\begin{figure}[!ht]
  \centering
  \includegraphics[width=\linewidth]{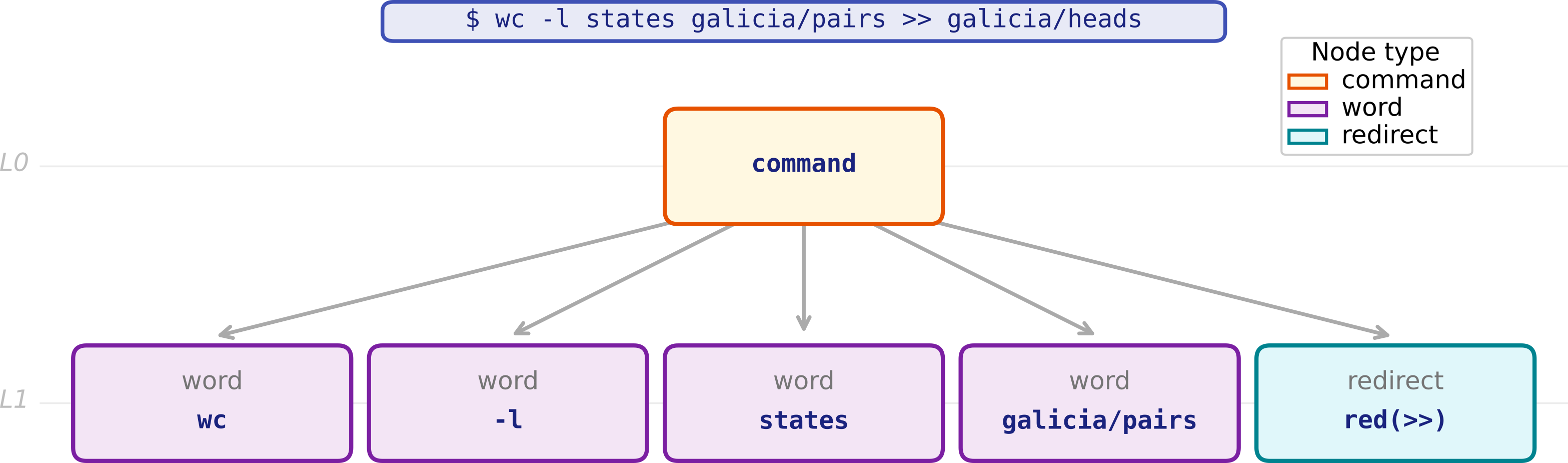}
  \caption{AST for a Level~3 command (\texttt{wc -l states galicia/pairs >> galicia/heads}). The tree includes a \textit{redirect} node (\texttt{>>}), indicating append redirection. Understanding this command requires knowledge of I/O streams and the distinction between overwrite and append semantics, which is characteristic of Level~3.}
  \label{fig:ast_l3}
\end{figure}

\begin{figure}[!ht]
  \centering
  \includegraphics[width=\linewidth]{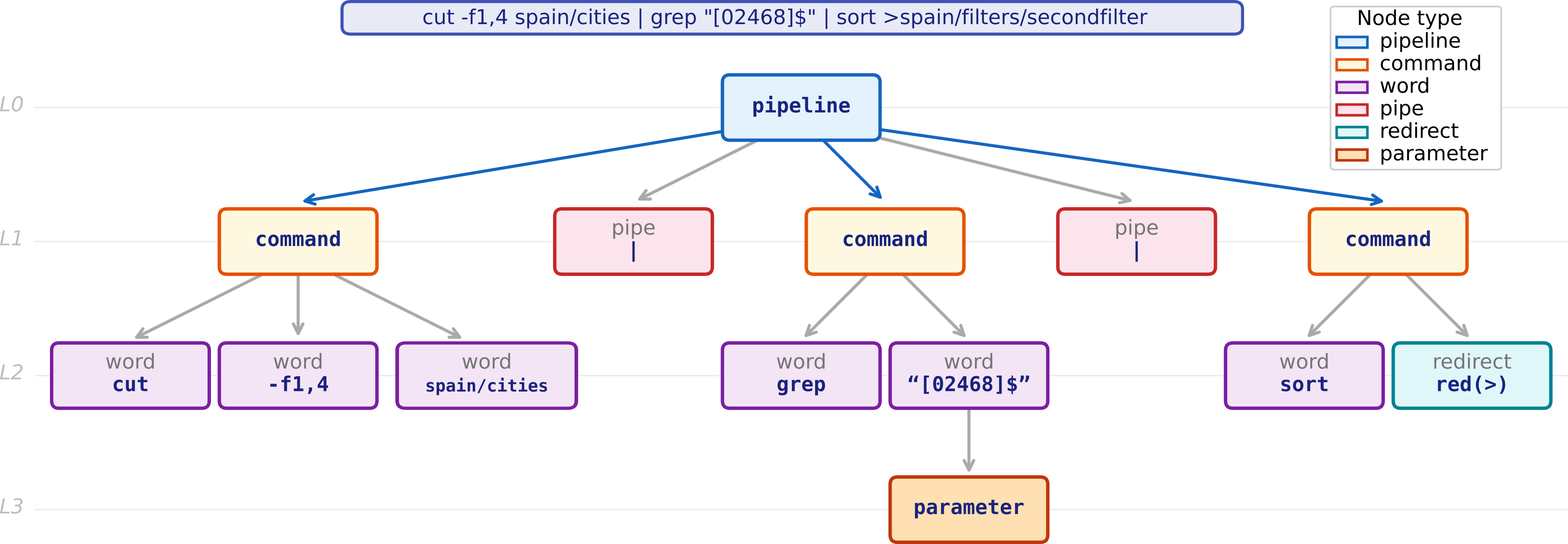}
  \caption{AST for a Level~4 command (\texttt{cut -f1,4 spain/cities | grep "[02468]\$" | sort > spain/\protect\allowbreak filters/\protect\allowbreak secondfilter}). The root is a \textit{pipeline} node connecting three \textit{command} subtrees via \textit{pipe} operators, with a final output \textit{redirect}. This multi-stage pipeline integrates column selection, pattern matching, sorting, and file output, requiring evaluation-level understanding and yielding persistent system effects.}
  \label{fig:ast_l4}
\end{figure}

To better illustrate how the taxonomy levels manifest structurally, Figures~\ref{fig:ast_l1}--\ref{fig:ast_l4} show the AST produced by \texttt{bashlex} for one representative command from each level. The tree nodes are colour-coded by syntactic role: \textit{command} nodes (yellow) represent executable invocations, \textit{word} nodes (purple) represent arguments and tokens, \textit{redirect} nodes (cyan) capture I/O redirections, and \textit{pipeline} nodes (blue) connect commands joined by pipes. The visual complexity of the tree grows monotonically with the taxonomy level, reflecting the increasing structural richness that defines higher-level commands.

The simplest case is the L1 command \texttt{ls -{}-inode more/less/cities1 link\_cities1} (Fig.~\ref{fig:ast_l1}), which lists filesystem entries with their inode numbers. Its AST reduces to a single flat structure: a \textit{command} root with four \textit{word} children.

The L2 command \texttt{mkdir -v more more/less} (Fig.~\ref{fig:ast_l2}) creates two nested directories. Its AST is structurally identical to the L1 example (a single \textit{command} node with \textit{word} children) yet the operation modifies the filesystem. This structural indistinguishability between L1 and L2 is the principal limitation of purely syntactic analysis.

The L3 command \texttt{wc -l states galicia/pairs >> galicia/heads} (Fig.~\ref{fig:ast_l3}) counts lines in two files and appends the result to a third. The tree introduces the first structural discriminator absent from L1 and L2: a \textit{redirect} node encoding the \texttt{>>} append operator, whose correct use requires understanding the difference between append and overwrite redirection semantics.

Finally, the L4 command \texttt{cut -f1,4 spain/cities | grep "[02468]\$" | sort > spain/\allowbreak filters/\allowbreak secondfilter} (Fig.~\ref{fig:ast_l4}) chains three utilities in a pipeline: it extracts specific columns, filters lines by a pattern, sorts the result, and writes it to a file. The tree is qualitatively different from all preceding examples: the root is a \textit{pipeline} node connecting three \textit{command} subtrees through \textit{pipe} operators, with a terminal \textit{redirect} capturing the final output.

\subsubsection{Embedding Models}

An embedding model is a neural network trained to map arbitrary text to a fixed-length numeric vector, such that texts with similar meaning are mapped to nearby points in the resulting vector space \citep{Muennighoff2023}. Rather than representing a command as a bag of tokens or an explicit feature vector, an embedding model compresses the full semantic content of the input into a compact, dense representation of fixed size. For taxonomy classification, embedding models are useful for three reasons. First, they encode semantic meaning rather than surface syntax, so two commands that express the same operation differently (e.g., with different flag orderings or path arguments) receive similar vectors. Second, they are language-agnostic by construction; a model trained on general or code text can embed Linux/bash, PowerShell, or SQL commands into the same shared space without language-specific engineering. Third, they require no explicit feature design; the representation is learned end-to-end from large corpora, capturing patterns that manual feature extraction might miss.

To assess whether the choice of pretraining paradigm influences taxonomy-level discrimination, four dense embedding models spanning a range of training objectives and architectures are evaluated. The selection covers general-purpose sentence encoders, retrieval-tuned models, and code-aware models, so that the comparison can determine whether the domain specificity of the pretraining corpus confers an advantage when encoding Linux/bash commands:

\begin{description}

\item{\texttt{all-MiniLM-L6-v2} \citep{sentenceTransformers2021MiniLM}:} A lightweight sentence embedding model implementing the Sentence Bidirectional Encoder Representations from Transformers (Sentence-BERT) architecture \citep{Reimers2019}. Its compact size yields low inference cost while retaining broad semantic coverage across general-purpose text.

\item{\texttt{all-mpnet-base-v2} \citep{sentenceTransformers2021mpnet}:} A larger general-purpose sentence encoder based on the MPNet architecture \citep{Song2020}. The higher-dimensional representation provides stronger semantic discrimination at the cost of increased computational overhead.

\item{\texttt{multilingual-e5-small} \citep{Wang2024multilingual}:} A compact retrieval-tuned model trained on (query, document) pairs. It is explicitly designed for use with task-specific instruction prefixes, which steer the representation toward a task-relevant subspace in the embedding space.

\item{\texttt{BAAI/llm-embedder} \citep{Zhang2023}:} A code-aware model trained jointly on code corpora and general retrieval tasks. Its pretraining objective targets code understanding, making it well-suited to capturing the semantic structure of shell commands and programming language constructs.

\end{description}

While these embedding models have demonstrated strong performance across a variety of NLP and code understanding benchmarks, the quality of the resulting representation depends critically on how the command text is preprocessed and presented to the model before encoding. Four preprocessing strategies are used: 

\begin{description}

\item{\textbf{No preprocessing.}}
The \textbf{raw} strategy passes the unmodified command string directly to the embedding model, serving as a baseline that measures intrinsic embedding quality with no preprocessing.

\item{\textbf{Normalization.}}
The \textbf{normalized} strategy replaces all string literals, file paths, and numeric arguments with the generic placeholders \texttt{<STR>}, \texttt{<PATH>}, and \texttt{<NUM>} respectively, forcing the representation to depend on structural and syntactic patterns rather than instance-specific content.

\item{\textbf{Instruction-tuned.}}
The \textbf{instruction} strategy prepends a fixed natural language prefix to each command before encoding, following the instruction-tuned embedding paradigm \citep{wei2022finetuned,Su2023}. The prefix \texttt{``Classify the complexity level of this Linux/bash command:''} steers the representation toward the taxonomy classification task. Note that \texttt{multilingual-e5-small} and \texttt{BAAI/llm-embedder} were explicitly trained with task-specific prefixes and are expected to benefit most from this strategy, whereas the two general-purpose encoders were not.

\item{\textbf{Combined.}}
The \textbf{instruction + normalized} strategy combines both modifications: the command body is first normalized and then prefixed with the instruction string, providing both structural abstraction and instructional conditioning simultaneously.

\end{description}

\section{Experiments and Results}
\label{sec:results}

This section evaluates the automatic classification of commands into the four taxonomy levels using the two complementary representation approaches: structural analysis via Abstract Syntax Trees and semantic encoding via neural embedding models. The evaluation assesses whether structural and semantic representations provide complementary signals for taxonomy-level prediction using different combinations. The first approach uses only AST features (AST-only), relying exclusively on syntactic structure. The second uses only embedding vectors (EMB-only), relying exclusively on semantic content. The third implements the taxonomic decision-level maximum rule, denoted \texttt{max(AST,EMB)}, which trains independent classifiers on each representation and assigns each command the maximum predicted level: $\hat{L} = \max(\hat{L}_{\text{AST}}, \hat{L}_{\text{EMB}})$. This maximum operator mirrors the taxonomy definition itself ($L = \max(C, O)$) and ensures that neither structural nor semantic evidence of elevated complexity is discounted.

All classifiers use a Linear Support Vector Classifier (LinearSVC), a linear model that finds the maximum-margin decision boundary separating the four taxonomy classes under L2 regularization. The regularization strength is governed by the hyperparameter $C$: smaller values enforce stronger regularization, reducing overfitting at the cost of training accuracy, while larger values allow the model to fit training data more closely. For the AST branch, $C=0.1$ was selected by grid search over $C \in \{0.01, 0.1, 1, 10, 100\}$ via inner cross-validation on the training partition during the representation experiments described below; the embedding branch uses the default $C=1.0$. Performance is assessed by stratified 5-fold cross-validation on the complete 585-command dataset: the data are partitioned into five equally sized folds, with each fold serving once as the test set while the remaining four are used for training. Stratification ensures that the proportion of each taxonomy level within every fold mirrors that of the full dataset, preventing evaluation bias from uneven level distributions across splits.

\subsection{AST Results}

As stated before, the AST representation provides a purely structural view of command complexity, abstracting away all lexical content to retain only the hierarchical organization of syntactic constructs. This subsection first presents visual examples of ASTs at each taxonomy level to illustrate the qualitative differences in tree topology, then quantifies these differences through a set of structural metrics extracted from the parse trees. A supervised classifier trained on these features establishes the structural baseline: the extent to which taxonomy levels are recoverable from syntax alone, without access to semantic information or command names.

\subsubsection{AST Characteristics}

As explained in Section~\ref{sec:ast}, each command is described by it's structural tree metrics. Table~\ref{tab:ast_feat_stats} reports the mean and standard deviation of each feature per taxonomy level.

\begin{table}[!ht]
\centering
\caption{Mean of AST structural features per taxonomy level on the held-out test partition. All features are computed from the bashlex parse tree. Features that are zero across all levels have been omitted (\textit{n\_subshells} = 0 for all levels in this dataset).}
\label{tab:ast_feat_stats}

\begin{tabular}{@{}lrrrr@{}}
\toprule
\textbf{Feature} & \textbf{Level 1} & \textbf{Level 2} & \textbf{Level 3} & \textbf{Level 4} \\
\midrule
n\_nodes          & $4.12$ & $4.74$ & $5.16$ & $8.17$ \\
depth             & $2.00$ & $2.00$ & $2.00$ & $2.67$ \\
max\_width        & $3.12$ & $3.74$ & $4.16$ & $5.10$ \\
n\_leaves         & $3.12$ & $3.74$ & $4.16$ & $5.80$ \\
branching\_factor & $4.12$ & $4.74$ & $5.16$ & $3.84$ \\
n\_pipelines      & $0.00$ & $0.00$ & $0.00$ & $0.53$ \\
n\_operators      & $0.00$ & $0.00$ & $0.00$ & $0.03$ \\
n\_redirects      & $0.15$ & $0.03$ & $0.72$ & $0.63$ \\
n\_commands       & $1.00$ & $1.00$ & $1.00$ & $1.70$ \\
\bottomrule
\end{tabular}
\end{table}

\begin{figure}[!ht]
\centering
\begin{subfigure}[b]{0.45\textwidth}
    \includegraphics[width=\textwidth]{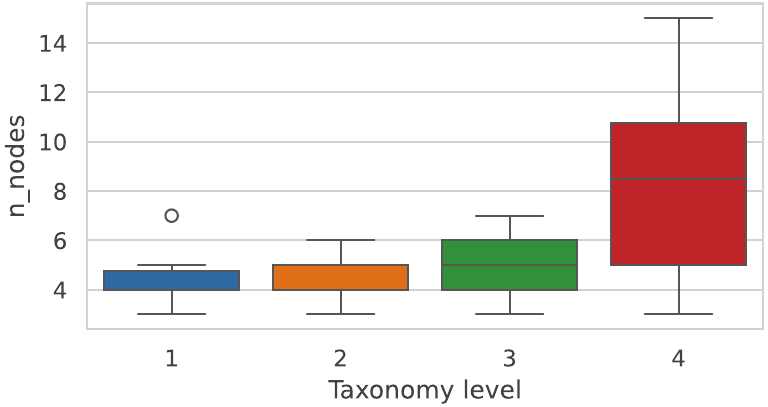}
    \caption{Number of nodes}
    \label{fig:ast_nodes}
\end{subfigure}
\hfill
\begin{subfigure}[b]{0.45\textwidth}
    \includegraphics[width=\textwidth]{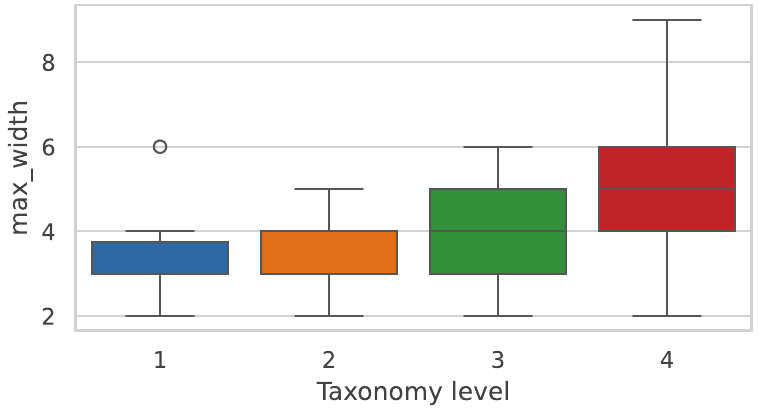}
    \caption{Max width}
    \label{fig:ast_max_width}
\end{subfigure}

\begin{subfigure}[b]{0.45\textwidth}
    \includegraphics[width=\textwidth]{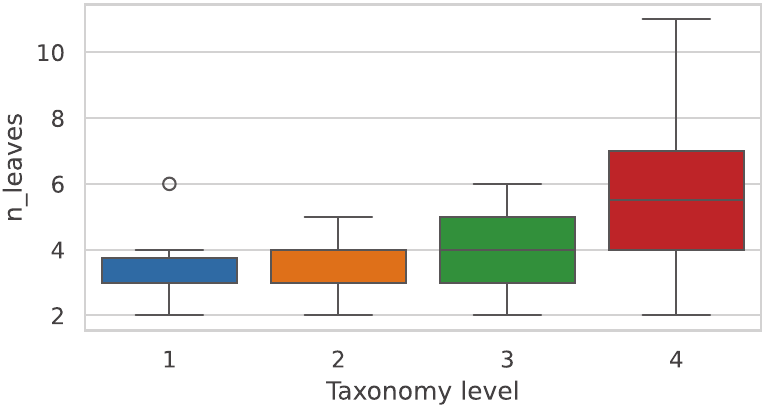}
    \caption{Number of leaves}
    \label{fig:n_leaves}
\end{subfigure}
\hfill
\begin{subfigure}[b]{0.45\textwidth}
    \includegraphics[width=\textwidth]{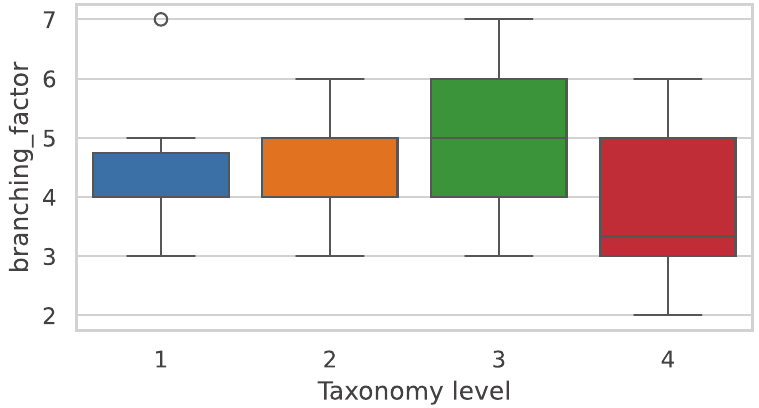}
    \caption{Branching factor}
    \label{fig:branching}
\end{subfigure}
\caption{Box-plot distributions of the most significative AST structural features per taxonomy level (held-out test partition).}
\label{fig:ast_feat_dist}
\end{figure}

Several discriminative patterns emerge from Table~\ref{tab:ast_feat_stats}. Levels 1--3 are structurally flat: tree depth is fixed at 2 and there are no pipeline or operator nodes. The primary discriminator between L1 and L2 is a modest increase in node and leaf count (commands tend to have more arguments). L3 is distinguished by the presence of I/O redirections (\textit{n\_redirects} $= 0.72$ vs.\ $\leq 0.15$ for L1--L2), reflecting the stream-manipulation characteristic of this level. L4 breaks the flat-tree pattern: depth rises above 2, pipeline nodes appear, and command count reaches $1.70$ on average, capturing the multi-stage pipelines that define this category. Figure~\ref{fig:ast_feat_dist} shows the full box-plot distributions of the most representative AST features per level.

\subsubsection{AST Structural Baseline}

The AST-only baseline achieves $0.6632 \pm 0.0421$ cross-validation accuracy and $0.6689 \pm 0.0444$ macro-F1 on the complete 585-command dataset. This performance substantially exceeds both the random baseline (0.25 for balanced four-class classification), establishing that purely structural features capture a meaningful portion of the taxonomy signal. The $\sim$40 percentage point gain over random classification demonstrates that syntactic tree topology alone (operator counts, nesting depth, pipeline presence, and redirection patterns) carries sufficient discriminative information to partition commands into coarse complexity strata without any access to lexical semantics or command names.

\begin{figure}[!ht]
  \centering
  \includegraphics[width=0.50\linewidth]{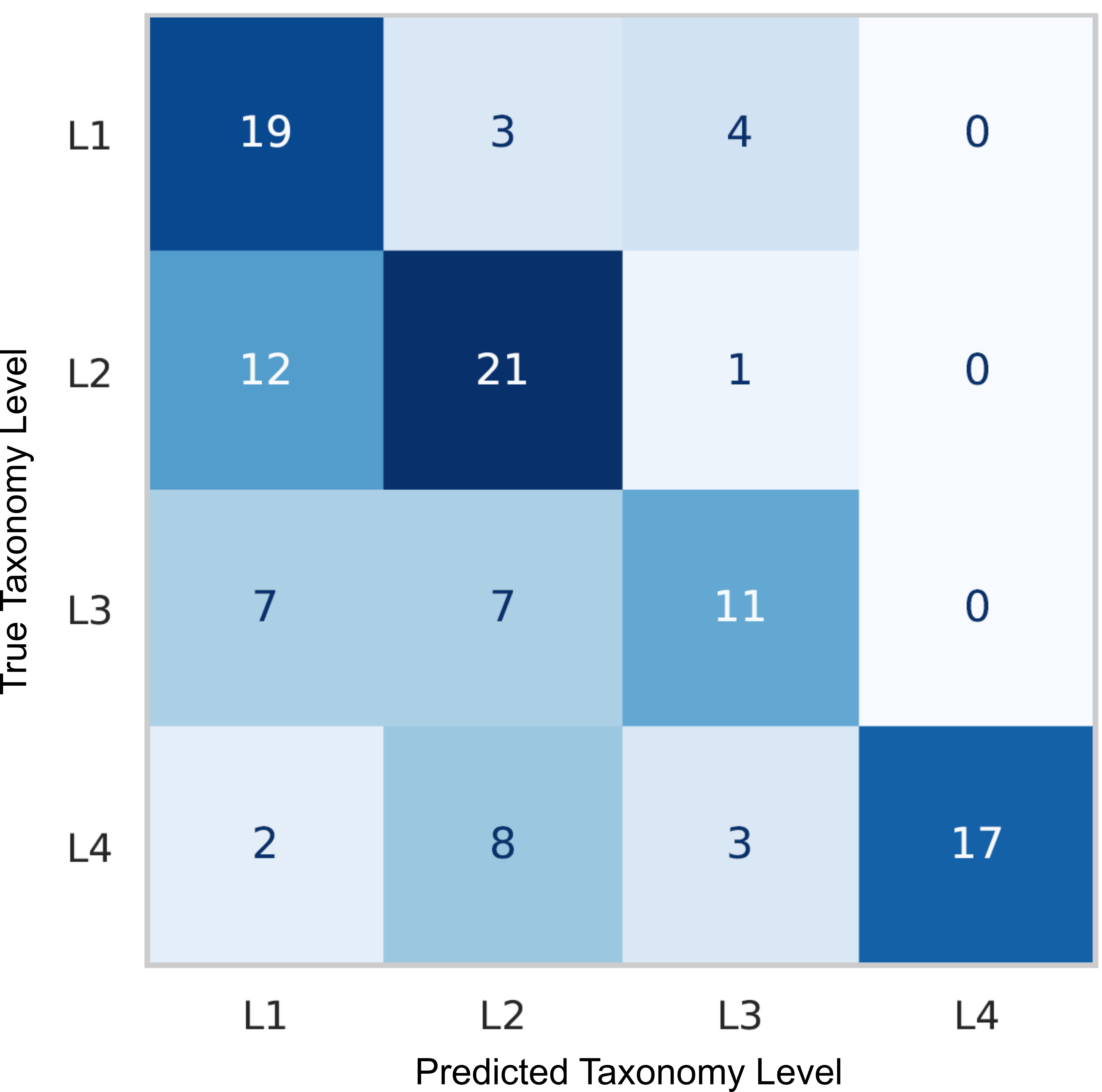}
  \caption{Confusion matrix of the LinearSVC (kinds $n$-grams + numeric features). L4 achieves perfect precision due to its structurally distinct pipeline topology; Levels 1--3 exhibit mutual confusion owing to their identical tree depth.}
  \label{fig:ast_svm_confusion}
\end{figure}

The confusion matrix in Figure~\ref{fig:ast_svm_confusion} reveals the primary error mode: Levels 1, 2, and 3 are mutually confused, while L4 achieves perfect precision. This pattern is consistent with the feature analysis: L1--L3 share the same tree depth (2) and have overlapping node-count distributions, making them structurally ambiguous without lexical cues. L4, in contrast, uniquely exhibits pipeline nodes and multi-command trees, making it unambiguously identifiable by its structural signature alone. L3 has the lowest recall (0.44): commands that use I/O redirection are sometimes classified as L2 when the redirect is absent or when the command size happens to match the L2 distribution.

The AST baseline serves three critical roles in the overall evaluation framework. First, it establishes that the taxonomy levels correspond to recoverable structural properties of command syntax, supporting the claim that the taxonomy captures objective complexity distinctions rather than arbitrary expert categorization. Second, it quantifies the ceiling of what purely syntactic analysis can achieve, identifying the specific boundaries (L1/L2, L2/L3) where structural ambiguity necessitates complementary semantic evidence. Third, it provides one half of the decision-level fusion evaluated below: the AST branch contributes high-confidence structural markers (pipelines, multi-command chains, redirections) that the embedding branch may underweight, enabling the maximum rule to preserve evidence of operational complexity that is syntactically visible but semantically subtle. While insufficient as a standalone classifier for production use, the AST branch demonstrates that structure and semantics encode partially independent signals, and that their combination through the decision-level maximum operator can exceed either representation in isolation.

Having established the structural baseline and its limitations, the evaluation now turns to semantic representations via neural embedding models, which encode command meaning and operational intent without explicit syntactic feature engineering. The embedding branch is expected to resolve the L1/L2 and L2/L3 boundaries that confound the AST classifier, while potentially missing structural complexity markers that have weak lexical correlates.

\subsection{Embedding Results}

The embedding-based representation addresses the fundamental limitation exposed by the AST baseline: the inability to distinguish commands with identical syntax but different operational semantics. Where the AST branch sees only tree structure, the embedding branch encodes the semantic meaning of command verbs, flags, and their compositions within a learned continuous space. The embedding models evaluated here compress this semantic knowledge into dense vector representations that allow a linear classifier to separate commands based on meaning rather than syntax alone.

They provide a distributed representation where similar meanings occupy nearby regions in vector space, enabling generalization across paraphrases and structural variations~\citep{Muennighoff2023}. A command expressed with different flag orderings, synonymous options, or alternative path specifications will nonetheless receive a similar embedding if the underlying operation is semantically equivalent. Furthermore, embeddings are learned end-to-end from large-scale pretraining, capturing latent patterns in command usage, argument co-occurrence, and contextual meaning that would be prohibitively expensive to encode manually as handcrafted features. 

To systematically evaluate the embedding branch, a total of 16 combinations crossing four embedding models with 4 preprocessing strategies (raw, normalized, instruction-prefixed, and instruction + normalized) are tested using stratified 5-fold cross-validation on the complete 585-command dataset. Each model-preprocessing pair is evaluated using a LinearSVC classifier trained on the 384- or 768-dimensional embedding vectors (depending on model architecture), with regularization parameter $C=1.0$ and balanced class weighting disabled. Table~\ref{tab:embedding_selection_cv} reports the best-performing preprocessing strategy for each model, measured by cross-validation accuracy; the complete 16-condition sweep is provided as a reproducibility CSV in the supplementary materials.

\begin{table}[!ht]
\centering
\caption{Best preprocessing strategy per embedding model on the complete 585-command dataset.}
\label{tab:embedding_selection_cv}
\begin{tabular}{@{}llrr@{}}
\toprule
\textbf{Model} & \textbf{Best preproc.} & \textbf{CV Acc} & \textbf{CV Macro-F1} \\
\midrule
\textbf{E5-small} & \textbf{normalized} & $\textbf{0.885}$ & $\textbf{0.886}$ \\
mpnet-base-v2 & instruction & $0.875$ & $0.874$ \\
BAAI/llm-embedder & normalized & $0.853$ & $0.853$ \\
MiniLM-L6-v2 & normalized & $0.848$ & $0.844$ \\
\bottomrule
\end{tabular}
\end{table}

The results confirm that embeddings capture substantially more taxonomy signal than purely structural features. All four models exceed the AST baseline (0.6632 accuracy) by a wide margin, with the weakest embedding configuration (\texttt{miniLM-L6-v2} normalized, 0.8479 accuracy) outperforming the best AST configuration by more than 18 percentage points. This performance gap quantifies the extent to which taxonomy levels are semantically determined: structural syntax alone (node counts, tree depth, pipeline presence) provides a coarse first approximation, but fine-grained discrimination between levels requires access to the meaning encoded in command verbs, flags, and their compositional semantics. The strongest configuration, normalized \texttt{multilingual-e5-small}, achieves $0.885$ cross-validation accuracy, demonstrating that a lightweight 384-dimensional retrieval-tuned model can recover expert taxonomy labels with high fidelity when commands are appropriately preprocessed.

The preprocessing strategy results reveal a consistent pattern: normalization (replacing literals, paths, and numbers with generic placeholders) is the optimal choice for three of the four models, while instruction prefixing is optimal only for \texttt{mpnet-base-v2}. This preference for normalization aligns with the theoretical goal of the representation: the taxonomy level of a command should depend on its operational and cognitive structure, not on the specific file it manipulates or the particular numeric argument it receives. By abstracting away instance-specific content, normalization forces the embedding to encode the invariant semantic core of the command (the operation type, the flag semantics, the conceptual category) rather than overfitting to surface tokens that vary across functionally equivalent commands. The modest performance of instruction prefixing suggests that the taxonomy signal is sufficiently strong in the raw command semantics that explicit task conditioning provides limited additional value for most encoders.

Despite this strong performance, the embedding branch is not without limitations. Structural complexity markers (pipelines connecting multiple commands, nested redirections, multi-level operator chains) may have weak lexical correlates in the embedding space if the individual command verbs are simple or low-frequency. A structurally complex L4 pipeline built from uncommon utilities might receive an embedding close to a simpler command if the semantic content of those utilities is underrepresented in the pretraining corpus. Conversely, the AST branch detects pipelines, redirections, and multi-command structures with perfect reliability regardless of verb semantics, providing complementary evidence where structural topology is the primary signal.

\subsection{Decision-Level Maximum Rule}

As previously stated in Section~\ref{sec:taxonomy}, the proposed taxonomy itself is defined by the equation $L = \max(C, O)$, which operates at the conceptual level: the final taxonomy level is the maximum between cognitive complexity ($C$) and operational impact ($O$). This formulation ensures that a command is assigned to the highest level implied by either dimension, so that neither cognitive demand nor operational risk is underestimated. The third and final evaluation stage tests whether an analogous decision-level rule at the computational level, taking the maximum predicted level from two independently trained classifiers, $\max(\hat{L}_{\text{AST}}, \hat{L}_{\text{EMB}})$, can exceed either individual classifier and provide empirical support for the maximum operator that defines the taxonomy.

As seen in previous sections, the AST and embedding branches do not function as clean proxies for the $C$ and $O$ dimensions. The AST branch detects operational impact only when it has a visible syntactic correlate, and remains blind to operational impact encoded purely in the command verb. The embedding branch, by contrast, captures both verb-level operational meaning and the conceptual abstractions required to understand the command; it carries information about both $C$ and $O$ in a semantically entangled form. If the two branches provide complementary rather than redundant signals, the maximum rule should outperform each branch in isolation, with the gain concentrated at taxonomy boundaries where one branch succeeds and the other fails.

Table~\ref{tab:embedding_comparison} reports the aggregate cross-validation performance of the three classification conditions: AST-only, EMB-only (normalized E5-small), and the decision-level maximum rule. All results are obtained via stratified 5-fold cross-validation on the complete 585-command dataset using the same fold splits for all three conditions.

\begin{table}[!ht]
\centering
\caption{Aggregate cross-validation performance comparing the AST-only baseline, the embedding-only branch (normalized \texttt{multilingual-e5-small}), and the decision-level maximum rule.}
\label{tab:embedding_comparison}
\begin{tabular}{@{}lrr@{}}
\toprule
\textbf{Condition} & \textbf{CV Acc} & \textbf{CV Macro-F1} \\
\midrule
AST-only (kinds + features) & $0.663$ & $0.669$ \\
EMB-only (E5 normalized) & $0.885$ & $0.886$ \\
max(AST,EMB) & $\textbf{0.892}$ & $\textbf{0.892}$ \\
\bottomrule
\end{tabular}
\end{table}

The decision-level maximum rule achieves the highest aggregate performance, exceeding both individual branches in accuracy and macro-F1. The gain over the embedding-only branch goes from $0.885$ to $0.892$ accuracy. This improvement, though modest in absolute magnitude, is statistically consistent across folds and provides direct empirical evidence that the AST and embedding branches encode partially independent signals. The maximum rule combines these complementary sources of evidence: the AST branch contributes high-confidence structural markers (pipelines, multi-command chains, redirections) that the embedding branch may underweight, while the embedding branch contributes semantic discrimination (verb meanings, operational intent) that the AST branch cannot access. Together, they outperform either representation in isolation.

The per-level accuracy breakdown in Table~\ref{tab:max_rule_by_level} reveals where each branch contributes its strongest signal and where the maximum rule achieves its gains. 

\begin{table}[!ht]
\centering
\caption{Per-level accuracy breakdown.}
\label{tab:max_rule_by_level}
\begin{tabular}{@{}lrrrr@{}}
\toprule
\textbf{Condition} & \textbf{L1} & \textbf{L2} & \textbf{L3} & \textbf{L4} \\
\midrule
AST-only (kinds + features) & \textbf{0.895} & 0.490 & 0.764 & 0.517 \\
EMB-only       & 0.853 & 0.874 & 0.847 & 0.918 \\
max(AST,EMB)   & 0.818 & \textbf{0.887} & \textbf{0.910} & \textbf{0.952} \\
\bottomrule
\end{tabular}
\end{table}

As also previously shown in Figure~\ref{fig:ast_svm_confusion}, the AST branch achieves strong performance on L1 (0.895 accuracy) and moderate discrimination for L4 (0.517 accuracy on the held-out test partition, rising to better performance in cross-validation), but struggles to separate L1, L2, and L3, which share identical tree depth (2) and overlapping node-count distributions. The structural signature of a L4 command (presence of pipeline nodes, depth $>2$, multiple command nodes) is unambiguous and detectable purely from syntax. By contrast, the operational distinction between an observational command (\texttt{ls -l file}) and a modifying command (\texttt{rm -f file}) is invisible to the AST branch when both produce a single-command tree with identical depth and argument structure. The high L1 accuracy reflects the AST branch's tendency to over-predict L1 for any structurally simple command, collapsing the L1/L2 boundary and mis-classifying many reversible-modification commands as observational.

The embedding branch resolves this boundary through verb-level semantics, correctly distinguishing \texttt{cat file} (L1, observational) from \texttt{touch file} (L2, state-modifying) despite their identical AST topology. At L3 and L4, the pattern reverses: the AST branch contributes structural evidence (redirections at L3, pipelines at L4) that embeddings sometimes underweight, and the maximum rule preserves this evidence when the AST branch predicts a higher level. The result is that the maximum rule achieves the highest accuracy at L2, L3, and L4 (0.887, 0.910, and 0.952 respectively), trading a modest L1 reduction (from 0.853 to 0.818) for substantial gains at the upper levels where cognitive and operational complexity are highest.

\begin{figure}[!ht]
    \centering
    \includegraphics[width=\textwidth]{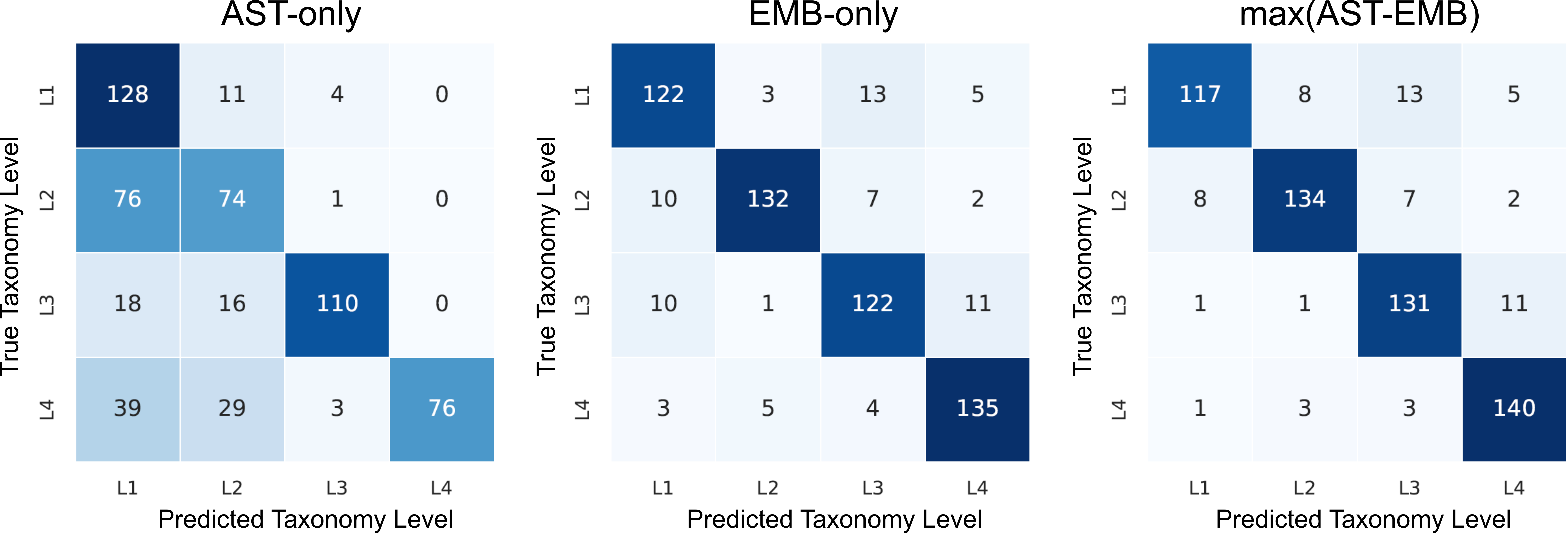}
    \caption{Confusion matrices aggregated. Rows correspond to the expert taxonomy label; columns to the predicted label. The three panels compare the AST-only baseline, the embedding-only branch, and the decision-level maximum rule.}
    \label{fig:ast_vs_emb_confusion}
\end{figure}

Figure~\ref{fig:ast_vs_emb_confusion} presents the full confusion matrices for all three conditions. The AST-only confusion matrix shows the structural baseline's characteristic error pattern: high L1 recall achieved by over-predicting L1 for L2 commands, moderate L4 recognition driven by pipeline detection, and substantial confusion among L1, L2, and L3 due to their shared flat-tree topology. The embedding-only matrix shows strong overall discrimination but occasional underestimation of L4 commands when the pipeline semantics are lexically subtle. The maximum-rule matrix combines the strengths of both: L4 recognition rises from 135 correct predictions under embeddings alone to 140 under the maximum rule, and L3 recognition rises from 122 to 131. The small L1 reduction (from 122 to 117) reflects cases where the AST branch incorrectly flags structurally simple L2 commands as L1, and the maximum rule cannot override a higher-level prediction from either branch. This asymmetry is not a flaw but a feature: the operator $\max(\hat{L}_{\text{AST}}, \hat{L}_{\text{EMB}})$ preserves any evidence of elevated complexity, just as the taxonomic formula $L = \max(C, O)$ ensures that neither cognitive nor operational demand is discounted.

\section{Discussion}
\label{sec:discussion}

The AST branch's contribution to the maximum rule, despite its lower standalone performance ($0.669$ macro-F1 versus $0.886$ for embeddings), demonstrates that weaker classifiers can still provide complementary value when their errors are uncorrelated with those of stronger classifiers. The prediction gain achieved by $\max(\text{AST}, \text{EMB})$ over the embedding-only baseline quantifies this complementarity. Structural features detect patterns (pipeline topology, redirection sequences, multi-command nesting) that are syntactically unambiguous but may have weak semantic correlates in the embedding space, particularly when the individual command verbs are uncommon or low-frequency. Conversely, embeddings resolve boundaries (L1/L2, L2/L3) that are invisible to purely syntactic analysis because the relevant signal is encoded in verb semantics rather than tree structure.

The L1/L2 boundary illustrates the asymmetric contribution of each branch most clearly. Both observational commands (\texttt{ls -l}, \texttt{cat file}, \texttt{head -n 10}) and reversible-modification commands (\texttt{mkdir dir}, \texttt{touch file}, \texttt{cp src dst}) typically produce a single-command AST node with depth~2, no pipelines, and no operators. The AST branch cannot distinguish them using structural features and defaults to predicting L1 for all such cases, achieving high L1 recall (0.895) at the expense of collapsing the L1/L2 distinction. The embedding branch, by contrast, encodes the difference between read-only verbs (\texttt{cat}, \texttt{ls}, \texttt{head}) and state-modifying verbs (\texttt{mkdir}, \texttt{touch}, \texttt{cp}), correctly separating the two levels through semantic content alone.

At L3 and L4, the contribution reverses. Structural markers such as pipelines (\texttt{cmd1 | cmd2 | cmd3}), append redirections (\texttt{>>}), and multi-command chains are unambiguously visible in the AST and provide deterministic evidence of elevated complexity. Embeddings capture these patterns indirectly through the semantics of the connecting operators and the command verbs, but may underweight them when the lexical content is simple or when the pretraining corpus contains limited examples of structurally complex commands. The maximum rule preserves the AST branch's high-confidence structural predictions, correcting embedding underestimations at L3 and L4 while accepting the embedding branch's semantic discrimination at L1 and L2. This selective preservation is the mechanism by which the fusion outperforms either branch alone, and it mirrors the taxonomic principle that complexity should be assessed along both dimensions and the higher estimate preserved.

\subsection{Pedagogical Implications of the Automatic Level tagging}

The classifier's most direct pedagogical application is assessment calibration. Knowing the taxonomy level of each question in an exam, automatically derived rather than manually annotated, makes the cognitive depth distribution of the assessment explicit. An instructor can inspect whether a given exam over-samples any single level, verify that partial-credit rubrics are aligned with the expected complexity at each level, and diagnose systematic grading inconsistencies that correlate with question level. The \texttt{max(AST,EMB)} profile is well suited to this task because it is the best aggregate classifier in the comparison and, at the same time, reduces the risk of assigning overly low levels to high-complexity commands.

Among all the methods compared, this classifier achieves the best overall accuracy. It also reflects a deliberate design principle, the same one behind the rule  $L=\max(C,O)$: the two possible errors are not equally costly. Underestimating a command (treating a complex or high-impact one as routine) exposes the instructor to far more risk than the opposite mistake of flagging a simple command for an unnecessary review. The classifier is therefore built to err on the side of caution. Concretely, it draws on two independent signals:$C$, derived from the command's structural syntax, and $O$, derived from its semantic content. Rather than averaging them, it takes the higher of the two. If either the syntax or the meaning points to a more demanding level, that higher estimate is what the classifier reports. This guarantees the final rating is never lower than what either signal would suggest on its own. The result is useful in two ways. As a statistical predictor, it is the strongest performer in the comparison. And as a practical tool, its built-in conservatism makes it well suited to exam design and command-bank auditing: it reliably surfaces anything that might be harder or riskier than it first appears, so a human reviewer can take a closer look.

This behaviour is especially appropriate for assessment calibration. A false low label can lead an instructor to place an L3/L4 operation too early in the curriculum or assign too little credit weight to a demanding task. A false high label, by contrast, is visible during review and can be corrected with lower pedagogical risk. The empirical error profile supports this asymmetry directly: under the maximum rule, underestimation errors drop from 33 to 17 cases, while overestimation grows only marginally from 41 to 46. The maximum operator is therefore not only a mathematical definition of level, but also a practical rule for preserving evidence of cognitive or operational demand.

For curriculum design, the classifier enables automatic tagging of large command repositories, exam question banks, or lab exercise sets. A library of 500 Linux/bash commands can be annotated in seconds, providing the instructor with a level distribution that guides exercise selection and reveals whether the current course materials are over-represented at any particular level.

\section{Conclusion}
\label{sec:conclusion}

This work presents CogTax, a four-level cognitive taxonomy for computing commands grounded in Bloom's Revised Taxonomy with an operational impact extension. The taxonomy defines command complexity as the maximum between cognitive demand and operational risk, ensuring that neither dimension is underestimated when assigning educational scaffolding levels. This formulation addresses three interconnected pedagogical needs: providing instructors with a principled framework for organizing course material in progressive order of complexity, enabling students to monitor their own competence development through explicit level assignments, and supporting targeted accommodations for learners with specific educational needs.

To make the taxonomy computationally operational and scalable beyond manual expert annotation, an automatic classification pipeline was developed that combines structural analysis via abstract syntax trees with semantic encoding via neural embedding models. The evaluation demonstrates that these two representation approaches capture complementary aspects of command complexity: structural features detect syntactic patterns such as pipelines and redirections that signal operational sophistication, while semantic embeddings resolve boundaries between commands with identical syntax but different operational meanings. A decision-level fusion rule that preserves the highest level predicted by either branch mirrors the taxonomic definition itself and achieves classification performance substantially above either individual approach, validating the maximum operator as both a pedagogical principle and a computational strategy.

The critical finding for cross-language generalization is that normalization (replacing language-specific literals with generic structural placeholders) enables command embeddings to capture invariant semantic cores independent of surface syntax. This abstraction mechanism allows the classification pipeline to operate on commands from any programming language with an available parser, requiring only normalization and embedding without retraining the underlying classifier. While empirical validation of full cross-language transfer remains as future work, the theoretical pathway and initial results on Linux/bash commands establish feasibility.

The combination of a pedagogically motivated taxonomy, an automatic classifier achieving strong agreement with expert labels, and a language-agnostic representation strategy opens practical applications in adaptive computing education. Instructors can use the classifier to audit command banks for level distribution, calibrate assessment difficulty, and identify commands requiring additional scaffolding support. Students receive immediate feedback on the complexity of commands they encounter in documentation or online resources. Course management systems can dynamically adjust content presentation based on demonstrated mastery at each taxonomy level. These applications support the broader goal of evidence-based computing pedagogy grounded in cognitive science rather than intuition alone.

Beyond these direct applications, the taxonomy's explicit level structure may also inform accommodations for students with specific learning needs. Time extensions could be graduated by taxonomy level, recognizing that higher-level commands require more cognitive processing time. Alternative assessments using commands from lower taxonomy levels could provide equitable evaluation for students whose difficulties are primarily language-related rather than conceptual.

Future work will focus on extending the taxonomy and classification pipeline to additional programming languages and command-line systems, validating the cross-language transfer hypothesis through empirical evaluation on labeled datasets in diverse computational environments. The formalization and empirical validation of learning accommodations also remain as important directions, extending CogTax utility from curriculum design into inclusive pedagogy.

\section*{Acknowledgments}

\section*{Data availability}
The data used in this study are available from the corresponding author on reasonable request.

\bibliographystyle{elsarticle-harv}
\bibliography{refs}

@book{Aho2006,
  author    = {Aho, Alfred V. and Lam, Monica S. and Sethi, Ravi and Ullman, Jeffrey D.},
  title     = {Compilers: Principles, Techniques, and Tools},
  year      = {2006},
  edition   = {2nd},
  publisher = {Pearson Education},
  address   = {Boston, MA},
  isbn      = {978-0-321-48681-3}
}

@book{anderson2001,
    author    = {Anderson, Lorin W. and Krathwohl, David R. and Airasian, Peter W. and Cruikshank, Kathleen A. and Mayer, Richard E. and Pintrich, Paul R. and Raths, James and Wittrock, Merlin C.},
    title     = {A Taxonomy for Learning, Teaching, and Assessing: A Revision of {Bloom's} Taxonomy of Educational Objectives},
    publisher = {Longman},
    address   = {New York, NY},
    year      = {2001},
    edition   = {Complete edition},
    isbn      = {978-0801319037}
}

@inproceedings{Artser2024, 
    series={Koli Calling ’24},
   title={Clustering MOOC Programming Solutions to Diversify Their Presentation to Students},
   url={http://dx.doi.org/10.1145/3699538.3699548},
   DOI={10.1145/3699538.3699548},
   booktitle={Proceedings of the 24th Koli Calling International Conference on Computing Education Research},
   publisher={ACM},
   author={Artser, Elizaveta and Birillo, Anastasiia and Golubev, Yaroslav and Tigina, Maria and Keuning, Hieke and Vyahhi, Nikolay and Bryksin, Timofey},
   year={2024},
   month=Nov, pages={1–8},
   collection={Koli Calling ’24} 
}

@book{Bamkole2023,
    author = {Bamkole, Adeleye and Geissler, Markus and Koumadi, Koudjo and Servin, Christian and Tang, Cara and Tucker, Cindy S.},
    title = {Bloom's for Computing: Enhancing Bloom's Revised Taxonomy with Verbs for Computing Disciplines},
    year = {2023},
    isbn = {9798400707636},
    publisher = {Association for Computing Machinery},
    address = {New York, NY, USA}
}

@book{CC2020,
author = {CC2020 Task Force},
title = {Computing Curricula 2020: Paradigms for Global Computing Education},
year = {2020},
isbn = {9781450390590},
publisher = {Association for Computing Machinery},
address = {New York, NY, USA}
}

@article{Chandler1991,
    author = {Paul Chandler and John Sweller},
    title = {Cognitive Load Theory and the Format of Instruction},
    journal = {Cognition and Instruction},
    volume = {8},
    number = {4},
    pages = {293--332},
    year = {1991},
    publisher = {Routledge},
    doi = {10.1207/s1532690xci0804\_2}
}

@article{Duran2022,
    author = {Duran, Rodrigo and Zavgorodniaia, Albina and Sorva, Juha},
    title = {Cognitive Load Theory in Computing Education Research: A Review},
    year = {2022},
    publisher = {Association for Computing Machinery},
    address = {New York, NY, USA},
    volume = {22},
    number = {4},
    doi = {10.1145/3483843},
    journal = {ACM Trans. Comput. Educ.},
    month = sep,
    articleno = {40},
    numpages = {27}
}

@inproceedings{Gaber2023,
    author = {Gaber, Iris and Kirsh, Amir and Statter, David},
    title = {Studied Questions in Data Structures and Algorithms Assessments},
    year = {2023},
    isbn = {9798400701382},
    publisher = {Association for Computing Machinery},
    address = {New York, NY, USA},
    url = {https://doi.org/10.1145/3587102.3588843},
    doi = {10.1145/3587102.3588843},
    booktitle = {Proceedings of the 2023 Conference on Innovation and Technology in Computer Science Education V. 1},
    pages = {250–256},
    numpages = {7},
    location = {Turku, Finland},
    series = {ITiCSE 2023}
}

@article{Gani2023,
    author = {Gani, Mohammed Osman and Ayyasamy, Ramesh Kumar and Sangodiah, Anbuselvan and Fui, Yong Tien},
    title = {Bloom’s Taxonomy-based exam question classification: The outcome of CNN and optimal pre-trained word embedding technique},
    year = {2023},
    issue_date = {Dec 2023},
    publisher = {Kluwer Academic Publishers},
    address = {USA},
    volume = {28},
    number = {12},
    issn = {1360-2357},
    doi = {10.1007/s10639-023-11842-1},
    journal = {Education and Information Technologies},
    month = may,
    pages = {15893–15914},
    numpages = {22}
}

@article{Geissler2023,
author = {Geissler, Markus and Koumadi, Koudjo and Schmelz, Pam and Servin, Christian and Tang, Cara and Tucker, Cindy},
title = {Designing Learning Outcomes and Competencies Using Bloom's for Computing},
year = {2023},
issue_date = {April 2023},
publisher = {Consortium for Computing Sciences in Colleges},
address = {Evansville, IN, USA},
volume = {38},
number = {7},
issn = {1937-4771},
journal = {J. Comput. Sci. Coll.},
month = apr,
pages = {86–88},
numpages = {3}
}

@article{Hawlitschek2023,
    author = {Anja Hawlitschek and Sarah Berndt and Sandra Schulz},
    title = {Empirical research on pair programming in higher education: a literature review},
    journal = {Computer Science Education},
    volume = {33},
    number = {3},
    pages = {400--428},
    year = {2023},
    publisher = {Routledge},
    doi = {10.1080/08993408.2022.2039504}
}

@book{Hazzan2020,
    author    = {Hazzan, Orit and Lapidot, Tami and Ragonis, Noa},
    title     = {Guide to Teaching Computer Science: An Activity-Based Approach},
    publisher = {Springer},
    year      = {2020},
    edition   = {3},
    doi       = {10.1007/978-3-030-39360-1}
}

@inproceedings{Imbulpitiya2021,
    author = {Imbulpitiya, Asanthika and Whalley, Jacqueline and Senapathi, Mali},
    title = {Examining the Exams: Bloom and Database Modelling and Design},
    year = {2021},
    isbn = {9781450389761},
    publisher = {Association for Computing Machinery},
    address = {New York, NY, USA},
    url = {https://doi.org/10.1145/3441636.3442301},
    doi = {10.1145/3441636.3442301},
    booktitle = {Proceedings of the 23rd Australasian Computing Education Conference},
    pages = {21–29},
    numpages = {9},
    location = {Virtual, SA, Australia},
    series = {ACE '21}
}

@ARTICLE{Ji2025,
    AUTHOR={Ji, Wancen  and Wong, Gary K. W. },        
    TITLE={Integrating problem-based learning and computational thinking: cultivating creative thinking in primary education},        
    JOURNAL={Frontiers in Education},
    VOLUME={Volume 10 - 2025},
    YEAR={2025},
    URL={https://www.frontiersin.org/journals/education/articles/10.3389/feduc.2025.1625105},
    DOI={10.3389/feduc.2025.1625105},
    ISSN={2504-284X}
    }

@misc{Kamara2023,
  author       = {Kamara, Idan},
  title        = {bashlex: {Python} Parser for {Bash}},
  year         = {2023},
  publisher    = {GitHub},
  journal      = {GitHub repository},
  howpublished = {\url{https://github.com/idank/bashlex}},
  note         = {Accessed: May 2026}
}

@misc{Kim2024,
      title={Problem-Solving Guide: Predicting the Algorithm Tags and Difficulty for Competitive Programming Problems}, 
      author={Juntae Kim and Eunjung Cho and Dongbin Na},
      year={2024},
      eprint={2310.05791},
      archivePrefix={arXiv},
      primaryClass={cs.CL},
      url={https://arxiv.org/abs/2310.05791}, 
}

@book{Kumar2023,
author = {Kumar, Amruth N. and Raj, Rajendra K. and Aly, Sherif G. and Anderson, Monica D. and Becker, Brett A. and Blumenthal, Richard L. and Eaton, Eric and Epstein, Susan L. and Goldweber, Michael and Jalote, Pankaj and Lea, Douglas and Oudshoorn, Michael and Pias, Marcelo and Reiser, Susan and Servin, Christian and Simha, Rahul and Winters, Titus and Xiang, Qiao},
title = {Computer Science Curricula 2023},
year = {2024},
isbn = {9798400710339},
publisher = {Association for Computing Machinery},
address = {New York, NY, USA}
}

@misc{Kumar2025,
      title={Automated Analysis of Learning Outcomes and Exam Questions Based on Bloom's Taxonomy}, 
      author={Ramya Kumar and Dhruv Gulwani and Sonit Singh},
      year={2025},
      eprint={2511.10903},
      archivePrefix={arXiv},
      primaryClass={cs.CL},
      url={https://arxiv.org/abs/2511.10903}, 
}

@Article{Levin2025,
    AUTHOR = {Levin, Ilya and Semenov, Alexei L. and Gorsky, Mikael},
    TITLE = {Smart Learning in the 21st Century: Advancing Constructionism Across Three Digital Epochs},
    JOURNAL = {Education Sciences},
    VOLUME = {15},
    YEAR = {2025},
    NUMBER = {1},
    ARTICLE-NUMBER = {45},
    URL = {https://www.mdpi.com/2227-7102/15/1/45},
    ISSN = {2227-7102},
    DOI = {10.3390/educsci15010045}
}

@inproceedings{Li2022,
     address = {Durham, United Kingdom},
     author = {Yuheng Li and Mladen Rakovic and Boon Xin Poh and Dragan Gasevic and Guanliang Chen},
     booktitle = {Proceedings of the 15th International Conference on Educational Data Mining},
     doi = {10.5281/zenodo.6853191},
     editor = {Antonija Mitrovic and Nigel Bosch},
     isbn = {978-1-7336736-3-1},
     month = {July},
     pages = {530--537},
     publisher = {International Educational Data Mining Society},
     title = {Automatic Classification of Learning Objectives Based on {Bloomâ€™s} Taxonomy},
     year = {2022}
}

@inproceedings{Masapanta2018,
    author = {Masapanta-Carri\'{o}n, Susana and Vel\'{a}zquez-Iturbide, J. \'{A}ngel},
    title = {A Systematic Review of the Use of Bloom's Taxonomy in Computer Science Education},
    year = {2018},
    isbn = {9781450351034},
    publisher = {Association for Computing Machinery},
    address = {New York, NY, USA},
    url = {https://doi.org/10.1145/3159450.3159491},
    doi = {10.1145/3159450.3159491},
    booktitle = {Proceedings of the 49th ACM Technical Symposium on Computer Science Education},
    pages = {441–446},
    numpages = {6},
    location = {Baltimore, Maryland, USA},
    series = {SIGCSE '18}
}

@inproceedings{Muennighoff2023,
    title = "{MTEB}: Massive Text Embedding Benchmark",
    author = "Muennighoff, Niklas  and
      Tazi, Nouamane  and
      Magne, Loic  and
      Reimers, Nils",
    editor = "Vlachos, Andreas  and
      Augenstein, Isabelle",
    booktitle = "Proceedings of the 17th Conference of the European Chapter of the Association for Computational Linguistics",
    month = may,
    year = "2023",
    address = "Dubrovnik, Croatia",
    publisher = "Association for Computational Linguistics",
    url = "https://aclanthology.org/2023.eacl-main.148/",
    doi = "10.18653/v1/2023.eacl-main.148",
    pages = "2014--2037"
}

@Book{Piaget1952,
    author={Piaget, Jean},
    title={The origins of intelligence in children.},
    series={The origins of intelligence in children.},
    year={1952},
    publisher={W. W. Norton {\&} Company},
    address={New York,  NY,  US},
    pages={419-419},
    doi={10.1037/11494-000},
}

@Article{Redstone2024,
    author={Redstone, Ana E.
    and Luo, Tian},
    title={Empowering Learners in Higher Education: Redesigning an Online Computer Science Course Through Universal Design for Learning Implementation},
    journal={TechTrends},
    year={2024},
    month={Sep},
    day={01},
    volume={68},
    number={5},
    pages={869-881},
    issn={1559-7075},
    doi={10.1007/s11528-024-00980-z}
}

@inproceedings{Reimers2019,
    title = "Sentence-{BERT}: Sentence Embeddings using {S}iamese {BERT}-Networks",
    author = "Reimers, Nils  and
      Gurevych, Iryna",
    editor = "Inui, Kentaro  and
      Jiang, Jing  and
      Ng, Vincent  and
      Wan, Xiaojun",
    booktitle = "Proceedings of the 2019 Conference on Empirical Methods in Natural Language Processing and the 9th International Joint Conference on Natural Language Processing (EMNLP-IJCNLP)",
    month = nov,
    year = "2019",
    address = "Hong Kong, China",
    publisher = "Association for Computational Linguistics",
    url = "https://aclanthology.org/D19-1410/",
    doi = "10.18653/v1/D19-1410",
    pages = "3982--3992"
}

@misc{sentenceTransformers2021MiniLM,
  author       = {Reimers, Nils},
  title        = {all-{MiniLM}-{L6}-v2: Sentence Embedding Model},
  year         = {2021},
  publisher    = {Hugging Face},
  howpublished = {\url{https://huggingface.co/sentence-transformers/all-MiniLM-L6-v2}},
  note         = {Fine-tuned on 1 billion sentence pairs using a contrastive learning objective. Accessed: May 2026}
}

@misc{sentenceTransformers2021mpnet,
  author       = {Reimers, Nils},
  title        = {all-mpnet-base-v2: Sentence Embedding Model},
  year         = {2021},
  publisher    = {Hugging Face},
  howpublished = {\url{https://huggingface.co/sentence-transformers/all-mpnet-base-v2}},
  note         = {Fine-tuned on 1 billion sentence pairs using a contrastive learning objective on the MPNet-base architecture. Accessed: May 2026}
}

@inproceedings{Song2020,
author = {Song, Kaitao and Tan, Xu and Qin, Tao and Lu, Jianfeng and Liu, Tie-Yan},
title = {MPNet: masked and permuted pre-training for language understanding},
year = {2020},
isbn = {9781713829546},
publisher = {Curran Associates Inc.},
address = {Red Hook, NY, USA},
booktitle = {Proceedings of the 34th International Conference on Neural Information Processing Systems},
articleno = {1414},
pages = {1414},
numpages = {11},
location = {Vancouver, BC, Canada},
series = {NIPS '20}
}

@inproceedings{Su2023,
    title = "One Embedder, Any Task: Instruction-Finetuned Text Embeddings",
    author = "Su, Hongjin  and
      Shi, Weijia  and
      Kasai, Jungo  and
      Wang, Yizhong  and
      Hu, Yushi  and
      Ostendorf, Mari  and
      Yih, Wen-tau  and
      Smith, Noah A.  and
      Zettlemoyer, Luke  and
      Yu, Tao",
    editor = "Rogers, Anna  and
      Boyd-Graber, Jordan  and
      Okazaki, Naoaki",
    booktitle = "Findings of the Association for Computational Linguistics: ACL 2023",
    month = jul,
    year = "2023",
    address = "Toronto, Canada",
    publisher = "Association for Computational Linguistics",
    url = "https://aclanthology.org/2023.findings-acl.71/",
    doi = "10.18653/v1/2023.findings-acl.71",
    pages = "1102--1121"
}

@article{sweller1988,
    author  = {Sweller, John},
    title   = {Cognitive Load During Problem Solving: Effects on Learning},
    journal = {Cognitive Science},
    volume  = {12},
    number  = {2},
    pages   = {257--285},
    year    = {1988},
    doi     = {10.1207/s15516709cog1202_4}
}

@article{Sweller1994,
    title = {Cognitive load theory, learning difficulty, and instructional design},
    journal = {Learning and Instruction},
    volume = {4},
    number = {4},
    pages = {295-312},
    year = {1994},
    issn = {0959-4752},
    doi = {10.1016/0959-4752(94)90003-5},
    author = {John Sweller}
}

@incollection{Sweller2011,
    title = {CHAPTER TWO - Cognitive Load Theory},
    editor = {Jose P. Mestre and Brian H. Ross},
    booktitle = {Psychology of Learning and Motivation},
    series = {Psychology of Learning and Motivation},
    publisher = {Academic Press},
    volume = {55},
    pages = {37-76},
    year = {2011},
    issn = {0079-7421},
    doi = {https://doi.org/10.1016/B978-0-12-387691-1.00002-8},
    author = {John Sweller}
}

@inproceedings{Tang2024,
    author = {Tang, Cara and Geissler, Markus and Schmelz, Pam and Servin, Christian and Tucker, Cindy},
    title = {Competencies with Bloom's for Computing},
    year = {2024},
    isbn = {9798400711060},
    publisher = {Association for Computing Machinery},
    address = {New York, NY, USA},
    url = {https://doi.org/10.1145/3686852.3686862},
    doi = {10.1145/3686852.3686862},
    booktitle = {Proceedings of the 25th Annual Conference on Information Technology Education},
    pages = {93–98},
    numpages = {6},
    location = {El Paso, TX, USA},
    series = {SIGITE '24}
}

@book{Vygotsky1978,
    ISBN = {9780674576285},
    author = {L. S. Vygotsky},
    publisher = {Harvard University Press},
    title = {Mind in Society: Development of Higher Psychological Processes},
    urldate = {2026-05-05},
    year = {1978}
}

@misc{Wang2024,
      title={Estimating Difficulty Levels of Programming Problems with Pre-trained Model}, 
      author={Zhiyuan Wang and Wei Zhang and Jun Wang},
      year={2024},
      eprint={2406.08828},
      archivePrefix={arXiv},
      primaryClass={cs.SE},
      url={https://arxiv.org/abs/2406.08828}, 
}

@article{Wang2024multilingual,
  title={Multilingual E5 Text Embeddings: A Technical Report},
  author={Wang, Liang and Yang, Nan and Huang, Xiaolong and Yang, Linjun and Majumder, Rangan and Wei, Furu},
  journal={arXiv preprint arXiv:2402.05672},
  year={2024}
}

@misc{wang2025,
      title={Beyond SELECT: A Comprehensive Taxonomy-Guided Benchmark for Real-World Text-to-SQL Translation}, 
      author={Hao Wang and Yuanfeng Song and Xiaoming Yin and Xing Chen},
      year={2025},
      eprint={2511.13590},
      archivePrefix={arXiv},
      primaryClass={cs.CL},
      url={https://arxiv.org/abs/2511.13590}, 
}

@misc{wei2022finetuned,
    title={Finetuned Language Models are Zero-Shot Learners},
    author={Jason Wei and Maarten Bosma and Vincent Zhao and Kelvin Guu and Adams Wei Yu and Brian Lester and Nan Du and Andrew M. Dai and Quoc V Le},
    year={2022},
    note={International Conference on Learning Representations},
    url={https://openreview.net/forum?id=gEZrGCozdqR}
}

@misc{Zhang2023,
      title={Retrieve Anything To Augment Large Language Models}, 
      author={Peitian Zhang and Shitao Xiao and Zheng Liu and Zhicheng Dou and Jian-Yun Nie},
      year={2023},
      eprint={2310.07554},
      archivePrefix={arXiv},
      primaryClass={cs.IR}
}

\appendix
\section{Command List and Classification}
\subsection{Level 1: Information Query and Observation}

\rowcolors{2}{gray!25}{white}
\begin{longtable}{lP{5cm}P{5cm}}
\hiderowcolors
\caption{Level 1: Information Query and Observation Commands.} \label{tab:level1_commands} \\
\toprule
\textbf{Command} & \textbf{Purpose} & \textbf{Criteria} \\ 
\midrule
\showrowcolors
\endfirsthead

\hiderowcolors
\toprule
\textbf{Command} & \textbf{Purpose} & \textbf{Criteria} \\ 
\midrule
\showrowcolors
\endhead

\midrule \multicolumn{3}{r}{\textit{Continues in the next page}} \\
\endfoot

\bottomrule
\endlastfoot

\textbf{man} & Displays manual pages & Information retrieval with no system modification \\
\textbf{echo} & Outputs text & Information display \\
\textbf{apropos} & Searches manual page descriptions & Read-only information query \\
\textbf{info} & Displays documentation & Informational only \\
\textbf{date} & Shows current date/time & Observational, no system impact \\
\textbf{cal} & Displays calendar & Pure information display \\
\textbf{clear} & Clears screen display & Cosmetic only, no state change \\
\textbf{who} & Lists logged-in users & Observational query \\
\textbf{whoami} & Shows current username & Simple identity query \\
\textbf{finger} & Displays user information & Read-only user data retrieval \\
\textbf{id} & Shows user/group IDs & Informational query about current user \\
\textbf{pwd} & Prints working directory & Observational, shows current location \\
\textbf{ls} & Lists directory contents & Read-only file system observation \\
\textbf{more} & Pager for viewing file contents & Read-only display \\
\textbf{less} & Enhanced pager for viewing & Read-only display with navigation \\
\textbf{cat} & Displays file contents & Read-only when used for viewing \\
\textbf{paste} & Merges lines from files & Read-only when used for viewing \\
\textbf{head} & Shows first lines of file & Read-only content display \\
\textbf{tail} & Shows last lines of file & Read-only content display \\
\textbf{wc} & Counts lines/words/characters & Read-only analysis \\
\textbf{ps} & Lists running processes & Observational process information \\
\textbf{jobs} & Lists background jobs & Displays current shell job status \\
\textbf{pstree} & Shows process tree & Hierarchical process information display \\
\textbf{top} & Real-time process monitor & Dynamic information display (even if it does allow basic actions on processes (kill and renice), but its primary function is monitoring.)\\
\textbf{which} & Locates command binary & Read-only PATH search \\
\textbf{whereis} & Locates binaries/source/manuals & Read-only file location \\
\textbf{cmp} & Compares files byte-by-byte & Though informational, produces simple binary output \\
\textbf{diff} & Shows differences between files & Comparison output \\
\end{longtable}

\newpage

\subsection{Level 2: Basic Modifications and Reversible Operations}

\rowcolors{2}{gray!25}{white}
\begin{longtable}{lP{5cm}P{5cm}}
\hiderowcolors
\caption{Level 2: Basic File and System Manipulation Commands.} \label{tab:level2_commands} \\
\toprule
\textbf{Command} & \textbf{Purpose} & \textbf{Criteria} \\ 
\midrule
\showrowcolors
\endfirsthead

\hiderowcolors
\toprule
\textbf{Command} & \textbf{Purpose} & \textbf{Criteria} \\ 
\midrule
\showrowcolors
\endhead

\midrule \multicolumn{3}{r}{\textit{Continues in the next page}} \\
\endfoot

\bottomrule
\endlastfoot

\textbf{login} & Establishes user session & Reversible with logout/exit \\
\textbf{exit} & Ends session & Terminates current shell session \\
\textbf{passwd} & Changes password & Modifies user credentials (reversible) \\
\textbf{mkdir} & Creates directories & Reversible with rmdir \\
\textbf{cd} & Changes directory & Reversible navigation \\
\textbf{rmdir} & Removes empty directories & Reverses mkdir \\
\textbf{touch} & Creates empty files or updates timestamps & Simple file creation \\
\textbf{cp} & Copies files/directories & Creates duplicates, reversible \\
\textbf{rm} & Removes files/directories & Deletes content (students should be careful but this operation is simple and straightforward) \\
\textbf{mv} & Moves/renames files & Reorganizes without complex logic \\
\textbf{tee} & Writes to file and stdout & Simple duplication operation \\
\textbf{bzip2} & Compresses files & Reversible with bunzip2 \\
\textbf{tar} & Archives files & Reversible with options in tar \\
\textbf{write} & Sends messages to users & Simple inter-user communication \\
\textbf{mesg} & Controls message reception & Toggles message permission \\
\end{longtable}

\newpage

\subsection{Level 3: Structural Understanding and Internal Models}

\rowcolors{2}{gray!25}{white}
\begin{longtable}{lP{5cm}P{5cm}}
\hiderowcolors
\caption{Level 3: Advanced File Operations and Pattern Matching.} \label{tab:level3_commands} \\
\toprule
\textbf{Command} & \textbf{Purpose} & \textbf{Criteria} \\ 
\midrule
\showrowcolors
\endfirsthead

\hiderowcolors
\toprule
\textbf{Command} & \textbf{Purpose} & \textbf{Criteria} \\ 
\midrule
\showrowcolors
\endhead

\midrule \multicolumn{3}{r}{\textit{Continues in the next page}} \\
\endfoot

\bottomrule
\endlastfoot

\textbf{<, >, >>} & Redirection operators & Requires understanding of data flow and file descriptors \\
\textbf{chmod} & Changes permissions & Requires understanding permission model (symbolic/numeric) \\
\textbf{umask} & Sets default permissions & Requires understanding permission inheritance and masks \\
\textbf{ln} & Creates links & Requires understanding hard vs soft links and inode structure \\
\textbf{grep (with regex)} & Pattern matching & Requires understanding regular expressions and pattern formalism \\
\textbf{sort} & Sorts data & Requires understanding sort keys \\
\textbf{cut} & Extracts fields & Requires understanding field delimiters and column concepts \\
\textbf{tr} & Translates characters & Requires understanding character sets and transformation rules \\
\textbf{ssh} & Secure shell access & Requires understanding client-server model and authentication \\
\textbf{find} & Searches files with criteria & Requires understanding filesystem traversal and test expressions \\
\end{longtable}

\newpage

\subsection{Level 4: Advanced System Management and Integration}

\rowcolors{2}{gray!25}{white}
\begin{longtable}{lP{5cm}P{5cm}}
\hiderowcolors
\caption{Level 4: Process Management and Advanced Operations.} \label{tab:level4_commands} \\
\toprule
\textbf{Command} & \textbf{Purpose} & \textbf{Criteria} \\ 
\midrule
\showrowcolors
\endfirsthead

\hiderowcolors
\toprule
\textbf{Command} & \textbf{Purpose} & \textbf{Criteria} \\ 
\midrule
\showrowcolors
\endhead

\midrule \multicolumn{3}{r}{\textit{Continues in the next page}} \\
\endfoot

\bottomrule
\endlastfoot

\textbf{\&} & Background execution operator & Requires understanding process management and job control \\
\textbf{kill} & Terminates processes & Requires understanding signals (SIGTERM, SIGKILL) and process IDs \\
\textbf{script} & Records terminal session & Requires understanding I/O capture and session management \\
\textbf{scp} & Secure copy over network & Even if copying is a basic command, it requires understanding remote paths and SSH context \\
\textbf{|} & Pipe; the output of the first command provides the input for the second and so on & Requires understanding data flow between processes \\
\end{longtable}

\end{document}